%% file: main.tex
\documentclass[compsoc,conference,a4paper,10pt,times]{IEEEtran}
\usepackage{cite}
\usepackage{amsmath,amssymb,amsfonts}
\usepackage{graphicx}
\usepackage{textcomp}
\usepackage{bmpsize}
\usepackage{xcolor}
\usepackage{lipsum}
\usepackage[colorlinks=true,urlcolor=black]{hyperref}
\usepackage{xspace}
\usepackage{subcaption}
\usepackage{multirow}
\usepackage{algpseudocode}
\usepackage{algorithm}
\usepackage{breqn}
\usepackage{url}
\usepackage{siunitx}
\usepackage{makecell}
\usepackage{footnotebackref}
\usepackage{booktabs}

\input{macros}

\begin{document}

\title{DKVE: Decentralized Key Validation for End-to-End Encrypted Messaging}

\author{\IEEEauthorblockN{1\textsuperscript{st} Subin Song}
\IEEEauthorblockA{\textit{Dept. of Computer Science and Engineering} \\
\textit{Seoul National University}\\
Seoul, South Korea \\
sbsong66@snu.ac.kr}
\and
\IEEEauthorblockN{2\textsuperscript{nd} Taekyoung (Ted) Kwon}
\IEEEauthorblockA{\textit{Dept. of Computer Science and Engineering} \\
\textit{Seoul National University}\\
Seoul, South Korea \\
tkkwon@snu.ac.kr}
}

\maketitle

\begin{abstract}
End-to-end encrypted messaging systems depend on authentic public key distribution to prevent man-in-the-middle (MitM) attacks. Current solutions present a stark trade-off: out-of-band (OOB) verification provides strong security but lacks scalability for large contact lists, while key transparency (KT) systems enable automated verification at high storage costs and operational complexity.

We propose \sys (Decentralized Key Validation for End-to-End Encrypted Messaging), a protocol that validates public keys through privacy-preserving cross-validation within users' social graphs. When obtaining a contact's public key from a key server, clients query mutual contacts to verify they hold the same key, combining Oblivious Pseudorandom Functions (OPRF) and Oblivious Key-Value Stores (OKVS) to preserve privacy of both queries and contact lists. \sys employs a Sequential Probability Ratio Test (SPRT) to aggregate responses and detect server misbehavior with user-configurable error bounds.

We evaluate \sys through simulations on real social network datasets, demonstrating \sys can detect MitM attacks with exceeding 97\% for strong-to-moderate-tie networks. The remaining 3\% of cases require validation through alternative methods such as KT and OOB verification.
Our proof-of-concept implementation confirms feasibility for background operation on commodity hardware, in terms of the latency and bandwidth.

As \sys can reduce the frequency of KT queries by two orders of magnitude, it enables fundamental architectural shifts: KT directories can migrate from fast but space-inefficient Merkle trees to space-efficient data structures like RSA accumulators.
While \sys cannot replace existing methods entirely---suffering from bootstrapping problems and degraded performance on weak-tie networks---it provides a practical complementary key validation mechanism, making secure messaging more deployable for billion-user systems.
\end{abstract}


\input{sections/1-introduction}
\input{sections/2-background}
\input{sections/3-assumptions}
\input{sections/4-system}
\input{sections/5-security-analysis}
\input{sections/6-evaluation}

\input{sections/7-discussion}

\input{sections/8-future-work}

\input{sections/9-related-work}

\input{sections/10-conclusion}

\section*{Acknowledgment}
The authors employed generative AI tools to edit the text and fix typos and grammatical mistakes.

\bibliographystyle{IEEEtran}
\bibliography{DKVE}

\appendices

\input{sections/appendix.tex}

\end{document}

%% file: macros.tex
\newcommand{\sys}{\textit{DKVE}\xspace}
\newcommand{\proc}[1]{\mbox{\textproc{#1}}}

\newcommand{\facebook}{\textit{Facebook}\xspace}
\newcommand{\pokec}{\textit{Pokec}\xspace}
\newcommand{\vk}{\textit{VKontakte}\xspace}
\newcommand{\signal}{\textit{Signal}\xspace}
\newcommand{\telegram}{\textit{Telegram}\xspace}
\newcommand{\whatsapp}{\textit{WhatsApp}\xspace}

\newcommand{\signalprotocol}{\textit{Signal protocol}\xspace}

\newcommand{\fbds}{\texttt{facebook}\xspace}
\newcommand{\pkds}{\texttt{pokec}\xspace}
\newcommand{\vkds}{\texttt{vk}\xspace}

\newcommand{\mumal}{\mu_{\mathsf{mal}}}
\newcommand{\sigmal}{\sigma_{\mathsf{mal}}}
\newcommand{\nsprt}{n_{\mathrm{SPRT}}}    

\newcommand{\Ematch}{E_{\mathsf{match}}}
\newcommand{\Emismatch}{E_{\mathsf{mismatch}}}
\newcommand{\Hhon}{H_{\mathsf{hon}}}
\newcommand{\Hmal}{H_{\mathsf{mal}}}

\newcommand{\parakeet}{\textit{Parakeet}\xspace}    
\newcommand{\akd}{\textit{AKD}\xspace}    

%% file: sections/1-introduction.tex
\section{Introduction}

End-to-end encrypted (E2EE) messengers---such as \signal \cite{SignalMessenger}, \whatsapp \cite{WhatsApp} and \telegram \cite{Telegram}---protect message contents by ensuring that only endpoints hold decryption keys. However, most systems still rely on a directory service to distribute users' public keys, creating a critical attack surface: a malicious server can distribute incorrect keys to enable man-in-the-middle (MitM) attacks. Current defenses face a significant practicality gap. Out-of-band (OOB) key verification methods (e.g., \signal's safety numbers) provide high security assurance but are impractical for users with large contact lists. In contrast, key transparency (KT) systems \cite{CONIKS,SEEMless,Parakeet,OPTIKS}  automate verification but impose substantial server-side complexity and storage overhead in real-world deployments. With user populations potentially reaching billions, these operational costs become prohibitive even for streamlined KT designs.

We propose \emph{\sys: Decentralized Key Validation for End-to-End Encrypted Messaging}, which statistically validates contacts' keys by leveraging users' existing social networks, designed to complement rather than replace OOB and KT approaches. The core insight is to bootstrap from a small number of high-confidence validations (obtained via OOB or KT) and propagate these assurances across mutual contacts. When a client retrieves a contact's key from the key server, it also privately queries selected mutual contacts to verify whether they possess the same key for the contact.
Suppose Alice and Bob both know Carol, Alice can compare the public key of Carol that she obtains (from the key server) with the one that Bob already has, and if they match, Alice can be more convinced that Carol's key from the server is correct.
In this process, we design \sys to preserve privacy by combining an Oblivious Pseudorandom Function (OPRF) with an Oblivious Key-Value Store (OKVS) \cite{OKVS}.

\sys adopts a \emph{reputation-aware} threat model: the key server may occasionally cheat and a fraction of a user's contacts may be malicious. Instead of absolute guarantees, \sys provides \emph{probabilistic detection} with user-set error bounds $(\alpha,\beta)$ on false positives and false negatives. The client aggregates match/mismatch evidences from responders (i.e., mutual contacts) with a Sequential Probability Ratio Test (SPRT) and stops as soon as the likelihood ratio crosses a decision threshold, thereby minimizing queries while meeting the desired error targets. This reduces the probability of falling back to OOB/KT to approximately in proportion to~$\alpha$ when the server is honest, while making server misbehaviors to be caught by $1-\beta$ probability. 
As the rate of incoming queries to the KT directory is substantially decreased, it can reduce the directory's online storage cost---at the expense of higher query processing costs---by moving the directory data to low-cost storage (say, from SSD to HDD) or by replacing the Merkle tree with a more space-efficient but query processing-inefficient data structure like RSA accumulators~\cite{kemmoeRSABasedDynamicAccumulator2024,baldimtsiObliviousAccumulators2024}.

We evaluate \sys through simulations on real social graph datasets from \facebook, \pokec, and \vk, implementing the system using the \textit{Matrix} E2EE messaging protocol. Our simulations quantify the typical number of responders required and analyze how attack detection rates correlate with the proportion of malicious contacts in the network. The prototype demonstrates that both computational and bandwidth overhead remain sufficiently low to enable background execution on commodity hardware.

\paragraph*{Contributions.} This paper makes four contributions:
\begin{enumerate}
    \item \textbf{Problem formulation.} We articulate the operational gaps in two existing methods for key authentication in E2EE messaging: (i) highly trustworthy but non-scalable OOB checks, and (ii) scalable but storage-heavy KT.
    \item \textbf{Protocol design.} We present \sys, a privacy-preserving cross-validation protocol that enables key comparison among users without revealing query targets or a responder's contact lists. We develop an SPRT-based decision procedure that achieves user-selected error bounds $(\alpha,\beta)$ while minimizing the number of queries. Furthermore, we analyze the protocol's security properties, including Sybil resistance and behavior under collusion.
    \item \textbf{Simulation.} We run simulations on real-world datasets, showing that \sys can successfully detect the server's MitM attacks with $\sim$ 97\% probability\footnote{Our evaluation yields approximately 97.2\% protocol success rate and 0.3\% false negative rate, resulting in $97.2\% \times (100-0.3)\% \approx 97\%$ detection rate.}, for strong-to-moderate-tie social graphs. Such high detection rate effectively deters the server from attempting to cheat, as the server's reputation is at stake.
    \item \textbf{Prototype implementation.} We implement a proof-of-concept (PoC) of \sys, demonstrating its practical latency (as background operations) and modest bandwidth usage.
\end{enumerate}

\noindent The remainder of this paper is organized as follows: \S\ref{sec:background} reviews the background on key authentication and the cryptographic tools we employ; \S\ref{sec:assumptions} formalizes the setting and goals; \S\ref{sec:system} details the \sys protocol and decision logic; \S\ref{sec:security-analysis} analyzes security properties; \S\ref{sec:evaluation} evaluates performance; \S\ref{sec:discussion} discusses practical considerations; \S\ref{sec:future-work} outlines future work; \S\ref{sec:related-work} reviews related work; and \S\ref{sec:conclusion} makes conclusive remarks.

%% file: sections/2-background.tex
\section{Background} \label{sec:background}
\subsection{End-to-End Encrypted (E2EE) Messaging}
In E2EE messaging services, such as \signal~\cite{SignalMessenger}, \whatsapp~\cite{WhatsApp}, and \telegram~\cite{Telegram}, messages are encrypted on the sender's device and decrypted exclusively on the receiver's device, ensuring that only the sender and receiver can read the messages. This design guarantees that even if the messaging service provider or its intermediary server is compromised, the message content remains private. The public keys of the sender and receiver are typically exchanged through a key directory server.

The \signalprotocol~\cite{SignalProtocol} is the most widely adopted protocol for E2EE messaging. It employs the X3DH (Extended Triple Diffie-Hellman) protocol~\cite{X3DH} for deriving a shared secret and the Double Ratchet algorithm for message encryption.
The X3DH key exchange provides session-level forward secrecy. Additionally, the sending and receiving keys are irreversibly updated (or ``ratcheted'') after each message, preventing key reuse and providing message-level forward secrecy. The \signalprotocol also incorporates the Diffie-Hellman (DH) ratchet algorithm to ensure post-compromise security. By combining these two ``ratcheting'' algorithms, \signal's post-handshake protocol is termed the ``Double Ratchet'' algorithm~\cite{DoubleRatchet}. The \signalprotocol has been adopted by numerous popular E2EE messaging services, including \whatsapp~\cite{WhatsApp}, \signal~\cite{SignalMessenger}, and \emph{Facebook Messenger}~\cite{FacebookMessenger, alatawiSoKAnalysisEndtoEnd2023}.

\subsection{Key Authentication in E2EE Messaging} \label{subsec:key-authentication}

\subsubsection{MitM Attack in E2EE Messaging}
When a key server managed by the messaging service provider handles the public keys of all users, the messaging system becomes vulnerable to man-in-the-middle (MitM) attacks. The key server may mistakenly or maliciously provide a sender with an incorrect public key for the intended receiver. Then, the sender unknowingly encrypts their message using the compromised key. If the key server intercepts this message, it can decrypt or tamper with the content using the corresponding private key.
Several key authentication methods, including out-of-band verification and key transparency, are currently employed to mitigate such MitM attacks in E2EE messaging services.

\subsubsection{Out-of-Band (OOB) Key Verification}
Popular E2EE messaging services often provide OOB key verification methods that allow users to manually verify the authenticity of their contacts' public keys.
In a \signal messenger~\cite{SignalMessenger}, users can verify their contacts' public keys by comparing ``safety numbers,'' which are generated by concatenating the hashes of both parties' $(\mathsf{username} \parallel \mathsf{public\_key})$ pairs, where the username corresponds to the user's phone number~\cite{moxiemarlinspikeSafetyNumberUpdates2016}.
The \signal mobile application allows users to scan QR codes containing safety numbers to verify that identical safety numbers appear on both devices. However, these out-of-band (OOB) methods suffer from scalability limitations since they require manual validation between each pair of users, making them impractical for verifying public keys across large contact networks.

\subsubsection{Key Transparency (KT)}
Key transparency (KT) is a system that enables users to cryptographically verify the authenticity and integrity of public keys retrieved from the key directory server in a scalable and automated manner.
KT systems such as \emph{CONIKS}~\cite{CONIKS}, \emph{SEEMless}~\cite{SEEMless}, and \parakeet~\cite{Parakeet} keep a server-side key directory that holds every user's public key together in the form of a cryptographic data structure (e.g., a Merkle tree or an append-only zero-knowledge set).
When a client looks up a contact's key, the server returns the key and a (non)membership proof, which the client verifies against the latest signed directory commitment. As commitments are cryptographically chained across epochs, users (and independent auditors) can verify that each new commitment extends the previous one---detecting tampering or omission.
To stop a malicious server from showing different views (or keys) to different users, these systems embed a global-consistency mechanism --- \emph{CONIKS} gossips tree-heads among clients and third-party auditors, while \parakeet employs a quorum of witnesses.

Several popular communication platforms have adopted KT systems to strengthen security. For instance, \whatsapp~\cite{WhatsAppKeyTransparency} and \emph{Proton Mail}~\cite{ProtonKeyTransparency} leverage KT systems to enable users to validate the authenticity of their contacts' public keys, thereby enhancing trust in their E2EE communications.

\subsection{OPRF and OKVS}

\subsubsection{Oblivious Pseudorandom Function (OPRF)} \label{subsubsec:oprf}

An Oblivious Pseudorandom Function (OPRF) is a cryptographic protocol that allows a client to evaluate a pseudorandom function (PRF) on an input without revealing the input to the server.

The inputs and outputs of a (batched) OPRF protocol are expressed as follows: 
\begin{itemize}
    \item \textbf{Client input:} The client holds a set of inputs $X = \{x_1, \ldots, x_n\}$.
    \item \textbf{Client output:} The client outputs a set of outputs $Y = \{y_1,\ldots, y_n\}$, where $y_i = PRF_k(x_i)$.
    \item \textbf{Server output:} The server outputs a random key $k$.
\end{itemize}

The following privacy properties hold:
\begin{itemize}
    \item \textbf{Server privacy:} The server learns nothing about the client's input set $X$, nor the output set $Y$.
    \item \textbf{Client privacy:} The client learns nothing about the secret key $k$.
\end{itemize}

\subsubsection{Oblivious Key-Value Store (OKVS)} \label{subsubsec:okvs}
An \emph{oblivious key-value store (OKVS)} is a data structure that encodes a set of label\footnote{To avoid confusion with cryptographic keys, we call OKVS keys as \emph{labels} in this paper.}-value pairs in such a way that an adversary, given only the encoding, learns nothing about the underlying labels or values. The term OKVS was first introduced in \cite{OKVS}. An OKVS consists of two core functions:
\begin{itemize}
    \item $\mathbf{S} \gets \mathsf{Encode}\left(\left\lbrace(l_1, v_1), \ldots, (l_n, v_n)\right\rbrace\right)$: The function takes a set of label-value pairs as input and outputs an encoded object $\mathbf{S}$, or $\bot$ with statistically small probability.
    \item $v \gets \mathsf{Decode}(\mathbf{S}, l)$: The function takes an encoded object $\mathbf{S}$ and a label $l$ as input, and outputs the value $v$ associated with the label $l$ in the encoded object.
\end{itemize}

\paragraph{Obliviousness}
When $L=\{l_1, \ldots l_n\}$ and $L'=\{l_1', \ldots l_n'\}$ are two different label sets (i.e., $L \neq L' $) and $V=\{v_1, \ldots v_n\}$ is a value set, the following two OKVS encodings are indistinguishable by an adversary:
\begin{displaymath}
    \mathbf{S} \gets \mathsf{Encode}(L, V), \quad \mathbf{S'} \gets \mathsf{Encode}(L', V)
\end{displaymath}

%% file: sections/3-assumptions.tex
\section{Assumptions} \label{sec:assumptions}

\subsection{Operating Environment}
We make the following assumptions about the operating environment of \sys. 
We assume \sys operates within an E2EE messaging service that relies on a \emph{centralized} relay server, such as \signal~\cite{SignalMessenger}. The same operator controls the message relay server and the key directory server.
Additionally, the messaging service provides out-of-band verification methods, such as \signal's safety numbers, alongside a KT service designed for infrequent queries.
The ``contact'' relationship is not necessarily symmetric; that is, Alice may have added Bob to her contacts, but Bob may not have added Alice to his.

\subsection{Threat Model} \label{subsec:threat-model}
\subsubsection{Key Directory Server} \label{subsubsec:threat-model-server}
The key directory server may be \emph{malicious} and attempt to provide incorrect public keys to users. However, since the server has a \emph{reputation} to maintain, it seeks to avoid launching overt or easily detectable attacks. Instead, it may attempt a \emph{man-in-the-middle (MitM) attack} that creates the illusion of legitimate communication between a user and their intended contact.

\subsubsection{Users}
The user who initiates a cross-validation request (to be detailed in \S\ref{subsec:cross-validation}) is called a ``querier'' and is \emph{honest but curious}. The querier performs the protocol correctly, as they seek to validate their contact's public key through the protocol. However, they may be curious and attempt to learn additional information about their contacts or other users. Another user who \emph{receives} the cross-validation request, called a ``responder,'' may be \emph{malicious} and does not care about their reputation. 
They may attempt to provide an incorrect public key to the querier or to learn additional information about the querier.

We assume that up to a fixed proportion of users in a given user's contact list, denoted by $\mumal$, are either \emph{malicious}, or \emph{compromised}. Both groups behave identically from the querier's perspective, as they provide a \emph{false public key}. Since they are indistinguishable in practice, we treat them as a single category and refer to them collectively as \emph{malicious users}.
Malicious users, potentially colluding with the server, may attempt to deceive queriers into accepting fraudulent keys as authentic. We assume that the proportion of malicious users does not exceed $\mumal$ with the practical validity of this assumption discussed in \S\ref{subsec:mumal-assumption}.

\subsection{Security Definition}

Leveraging the threat model in \S\ref{subsec:threat-model}, \sys employs a \emph{probabilistic approach} to detect server misbehaviors.
Once a user detects that the server may be cheating, they verify the information through an OOB method or a KT query and publicize the server's misbehavior online (on social networking services, online media, etc.). 

A \sys user defines $\alpha$ as the maximum acceptable false positive rate (the probability of falsely accusing an honest server of cheating)  and $\beta$ as  the maximum acceptable false negative rate (the probability of failing to detect a cheating server)\footnote{Note that $\alpha$ and $\beta$ can be configured as per-user or service-wide parameters.}. While \sys does not guarantee that a cheating server is ``always'' detected, it does so with a \emph{high probability} (greater than $1 -\beta$). If this probability is sufficiently large, \sys can effectively deter server misbehaviors, as the server would not want to risk its reputation by being caught cheating.

The security definition of \sys is that \emph{if the protocol succeeds, a cheating server is detected with the probability greater than $1-\beta$.} 
Detecting a malicious \emph{responder} is beyond the scope of this paper.

\subsection{Notation}

\subsection{Symbolic Notation}
Table~\ref{tab:notation} summarizes the symbolic notation used in this paper. Notation for algorithms is defined separately in Appendix~\ref{app:alg-notations}.
\input{tables/notation}

\subsection{Terminology}
To avoid the ambiguity between a user's public key and an OKVS key, we refer to an OKVS key as an \textit{``label''} rather than a \textit{``key''} throughout this paper.
Additionally, we abbreviate ``out-of-band verification'' as ``OOB verification'' and ``key transparency'' as ``KT.''

%% file: tables/notation.tex
\renewcommand{\arraystretch}{1.2}

\begin{table}[htbp]
    \centering
    \caption{Symbolic notation used throughout the paper.}
    \label{tab:notation}
    \begin{tabular}{ll}
      \toprule
      \textbf{Notation}                                                          & \textbf{Description}                                                                                                           \\ \midrule
      $\mathcal{V}$                                                                        & Set of all users in the system (vertices of a graph)                                                                           \\ \hline
      $\mathcal{E}$                                                                        & \begin{tabular}[c]{@{}l@{}}Set of all contact relationships in the system \\ (\emph{directed} edges of the graph)\end{tabular} \\ \hline
      $G := (\mathcal{V}, \mathcal{E})$                                                              & A \emph{directed} graph that represents the whole system                                                                       \\ \hline
      $(u, v) \in \mathcal{E}$                                                             & ``User $u$ has added user $v$ to their contact list''                                                                          \\ \hline
      ${\mathit{pk}}_u$                                                                     & Genuine public key of user $u$                                                                                                 \\ \hline
      $\widehat{\mathit{pk}}_{u}^{v}$                                      & \begin{tabular}[c]{@{}l@{}}Public key that the server claims to \\ user $v$ as user $u$'s public key\end{tabular}              \\ \hline
      $\mathcal{C}_u$                                                                      & User $u$'s contact list (set)                                                                                                  \\ \hline
      \begin{tabular}[c]{@{}l@{}}$q, r, t,$\\ $\mathcal{Q}, \mathcal{R}, \mathcal{T}$\end{tabular} & Querier, responder, target users (and their sets)                                                                              \\ \hline
      $\mumal$                                                                & Portion of malicious users in the system                                                                                       \\ \hline
      $\alpha, \beta$                                                            & Desired false positive/negative thresholds                                                                                     \\ \hline
      $\tau$                                                               & \begin{tabular}[c]{@{}l@{}}A fixed sequence of bytes (tag) that is assigned as a \\ prefix to public key bytes\end{tabular}          \\ 
      \hline
      $\parallel$ & Bytes concatenation operation \\
      \bottomrule
      \end{tabular}
  \end{table}

%% file: sections/4-system.tex
\section{The \sys System} \label{sec:system}
\subsection{Motivation}
Existing key authentication methods discussed in \S\ref{subsec:key-authentication} have the following limitations.
First, OOB verification such as \signal's safety number is not scalable, as it requires users to manually verify all contacts.
For key transparency (KT), server-side storage cost is widely recognized as a major challenge. \emph{CONIKS}~\cite{CONIKS} requires storage proportional to the number of server epochs, which is not scalable for real-world deployment.
While \emph{SEEMless}~\cite{SEEMless} and \parakeet~\cite{Parakeet} reduce the storage requirement to be proportional to the number of key updates, they require $\sim26.9~\mathrm{TB}$ of storage for 1 billion users (see Appendix~\ref{app:akd-storage-cost} for details).
\emph{OPTIKS}~\cite{OPTIKS} reduces the storage cost by half compared to \parakeet, but it still requires $\sim13.4~\mathrm{TB}$ of storage for 1 billion users.

\sys aims to serve as a \emph{complement} to OOB verification and KT to mitigate their scalability and storage cost limitations. The intuition is: \textbf{``bootstrap a small number of trusted points (i.e., public keys) using OOB and KT, and spread them across users through social graphs.''} By introducing \sys into E2EE messaging ecosystems, we can significantly reduce the frequency of KT queries, which either enables moving the entire KT directory to cheaper (and slower) storage or allows the use of more space-efficient data structures than Merkle trees. Additionally, the results of OOB verification---which is deemed most trustworthy but inconvenient---can be readily disseminated across multiple users via \sys.

\subsection{\sys Cross-Validation} \label{subsec:cross-validation}

\MakeRobust{\Call}

\sys provides a mechanism that allows users to \emph{cross-validate} the public key of their contact with another contact. This is feasible because users are likely to share mutual contacts within their social circles.
The purpose of \sys is to \emph{disseminate} verified keys (from OOB verification or KT) across users over their social networks, enabling other users to utilize those keys for cross-validation.

Key dissemination (or ``spreading'') occurs not only for keys verified via OOB and KT but also for keys that have been cross-validated via \sys.
This enables the keys validated by OOB/KT  to propagate across multiple hops in the social graph.
For instance, if Alice verifies Carol's key via OOB and Bob subsequently verifies Carol's key using \sys (with Alice as a responder), then Bob can further disseminate Carol's key to Dave via \sys. Even when the OOB/KT validation result serves as only one of multiple evidences in SPRT (to be discussed later), the influence of the OOB/KT validation can propagate throughout the social graph.

Key dissemination to another user occurs \emph{on-demand}, meaning keys are shared \textbf{only upon incoming key requests}.
Specifically, Bob provides Carol's key to Alice \emph{only upon} receiving a cross-validation request from Alice.
Consider the following scenario: Alice adds Carol to her contact list and retrieves Carol's public key, $\widehat{\mathit{pk}}_C^A$, from the key server.
Alice then initiates cross-validation requests to her contacts to verify Carol's public key. (The client application performs this process automatically without user intervention.)
Since Alice's client application cannot determine which contacts share Carol as a mutual contact, it randomly selects one of their contacts for cross-validation. Assume the selected contact is Bob.
A straightforward approach to cross-validate Carol's key with Bob would involve three steps: (1) Alice requests Carol's public key from Bob, (2) Bob responds with $\widehat{\mathit{pk}}_{C}^{B}$, and (3) Alice verifies that $\widehat{\mathit{pk}}_{C}^{A} = \widehat{\mathit{pk}}_{C}^{B}$. However, this approach compromises privacy, as Bob would learn which user's key Alice seeks to validate. To mitigate this privacy concern, \sys employs privacy-preserving cryptographic primitives: Oblivious Pseudorandom Functions (OPRFs) and Oblivious Key-Value Stores (OKVSs).

We refer to Alice, who initiates the cross-validation process, as the \emph{querier} $q$; Bob, who receives the cross-validation request, as the \emph{responder} $r$; and Carol, whose key is being validated, as the \emph{target} $t$ from now on.

The algorithm leveraging OPRF and OKVS is presented in Algorithm~\ref{alg:cross-validation-no-sprt} for the case in which $\mumal = 0$ (i.e., no malicious responders exist). The functions and notation used in Algorithm~\ref{alg:cross-validation-no-sprt} are summarized in Appendix~\ref{app:alg-notations}. The algorithm comprises two functions: \proc{CrossValQueryNaive}, which executes on the querier, and \proc{CrossValRespond}, which executes on the responder. \proc{CrossValQueryNaive} accepts as input the responder ($r$) and a set of targets for cross-validation ($\mathcal T$). Multiple targets are processed simultaneously rather than individually to leverage the efficiency of the batch OPRF protocol.
\input{algorithms/crossval-nosprt}

We initially assume that $\mumal = 0$, meaning all responders are honest. Under these conditions, the cross-validation process begins when the querier invokes \proc{CrossValQueryNaive}, and the responder receives the cross-validation request and executes \proc{CrossValRespond}.
The querier inputs the set of target usernames ($\mathcal T$) into the OPRF protocol. Both parties---querier and responder---then jointly execute the OPRF protocol. Consequently, the querier obtains the OPRF evaluations of the target usernames $\mathbf y$ (Line~\ref{line:query-oprfreceive}), while the responder obtains the PRF key $k$ (Line~\ref{line:respond-oprfsend}).
The responder subsequently applies the key $k$ to compute the PRF for each (plaintext) username in their contact list (Line~\ref{line:respond-oprfeval}).
Finally, the responder encodes their entire contact list into an OKVS object by calling \proc{OKVS.Encode}, using the OPRF evaluations of usernames as ``labels'' and the corresponding public keys as ``values'' (Line~\ref{line:respond-okvsencode}).

The responder sends the OKVS object to the querier (Line~\ref{line:respond-send}), who then decodes the object using the OPRF evaluations of target usernames as labels (Line~\ref{line:query-okvsdecode}). Since \proc{OKVS.Decode} produces seemingly random bytes even when decoded with a label absent from the responder's list, the querier cannot determine whether the responder actually possesses the target in their contact list from the output. To enable the querier to distinguish valid public keys from invalid ones (i.e., random bytes), the responder prefixes each public key with a fixed byte sequence $\tau$ (Line~\ref{line:respond-prefixadd}).
The querier verifies whether each decoded value begins with the fixed prefix $\tau$ (Line~\ref{line:check-prefix}). If the prefix is present, it is stripped, and the remaining portion is interpreted as the target's valid public key. (Otherwise, this indicates that the target is not in the responder's contact list.)
The querier then compares this public key with the one obtained from the server (Line~\ref{line:check-key-match}).
If the keys match, the querier updates the target's key status to \textsf{VALID}; otherwise, it is marked as \textsf{INVALID}.
This process is repeated for all targets in the set, and its completion signifies the end of \proc{CrossValQueryNaive}.

Due to the obliviousness properties of OPRF and OKVS, the following privacy guarantees hold in cross-validation:
\begin{itemize}
   \item \textbf{Querier privacy:} The responder learns no information about the querier's specific targets (beyond the query size) or the computed OPRF outputs, as these properties are guaranteed by OPRF obliviousness.
   \item \textbf{Responder privacy:} The querier gains no knowledge of the PRF key $k$ (protected by OPRF obliviousness) and learns nothing about the responder's contact list beyond the presence or absence of queried targets (ensured by OKVS obliviousness). However, the querier may be able to infer the size of the responder's contact list.
\end{itemize}

One may wonder why the OPRF evaluation (i.e., encrypted username) is used as a label in the OKVS encoding, rather than the plaintext username. The reason is to prevent a curious querier from performing a brute-force attack to discover the responder's entire contact list. If the responder were to use plaintext usernames as OKVS labels, the querier could attempt to brute-force the contact list by testing all possible usernames. Details regarding OKVS brute-force attack prevention are presented in \S\ref{subsec:okvs-brute-force}.

\subsection{How Many Peers to Query?} \label{subsec:crossval-sprt}

In the previous section, we assumed every responder is honest. However, as discussed in \S\ref{subsec:threat-model}, some responders may behave maliciously. Consequently, querying the same target across different responders can yield inconsistent responses. This raises critical questions: how many responses should the querier collect from responders, and how should they determine whether the server-provided key is valid when responses conflict?

To address this challenge, we employ the \emph{Sequential Probability Ratio Test (SPRT)}. SPRT is based on the likelihood ratio test, a well-established statistical method for hypothesis testing. The goal is to maintain both the false positive and false negative rates below predefined thresholds while minimizing the number of required queries. The SPRT algorithm is detailed in Appendix~\ref{app:sprt}.

Assume the querier $q$ wishes to validate $\widehat{\mathit{pk}}_{t}^{q}$ (from the key server). Let $\Hhon$ denote the hypothesis that $\widehat{\mathit{pk}}_{t}^{q} = \mathit{pk}_{t}$, i.e., the server-provided key is correct and the server is \emph{honest}. Let $\Hmal$ denote the hypothesis that $\widehat{\mathit{pk}}_{t}^{q} \neq \mathit{pk}_{t}$, i.e., the key is incorrect and the server is \emph{malicious}.
Let $\Ematch$ and $\Emismatch$ represent the evidence that $\widehat{\mathit{pk}}_{t}^{q}$ matches or does not match the key obtained from the \textproc{OKVS.Decode} result, respectively. Let $\alpha$ and $\beta$ denote the predefined thresholds of false positive and false negative rates. The decision boundaries $a$ and $b$ for SPRT are computed as:
\begin{equation}
    a = \log{\frac{\beta}{1 - \alpha}}, \quad b = \log{\frac{1 - \beta}{\alpha}}
\end{equation}

Each iteration $i$ (i.e., $i$-th responder) of SPRT proceeds as follows. The initial cumulative log-likelihood ratio (or ``SLLR'') $\Sigma_0$ is set to 0. The querier repeats the process below until a decision is reached or no more contacts remain to query:
\begin{enumerate}
    \item The querier collects evidence $E_i \in \{\Ematch, \Emismatch\}$.
    \item The querier updates the cumulative log-likelihood ratio $\Sigma_i$ as:
    \begin{equation}
        \Sigma_i = \Sigma_{i - 1} + \log{\frac{\Pr(E_i \mid \Hmal)}{\Pr(E_i \mid \Hhon)}}
    \end{equation}
    \item If $\Sigma_i \geq b$, the querier stops and concludes that the server is malicious. If $\Sigma_i \leq a$, the querier stops and concludes that the server is honest. Otherwise (i.e., $a < \Sigma_i < b$), the process continues to the next iteration.
\end{enumerate}

Note that $\Pr(\Ematch \mid \Hhon)$, $\Pr(\Ematch \mid \Hmal)$, $\Pr(\Emismatch \mid \Hhon)$, and $\Pr(\Emismatch \mid \Hmal)$ can be estimated as follows:
\begin{equation}
    \begin{split}
    \Pr(\Ematch \mid \Hhon) = \Pr(\Emismatch \mid \Hmal) = 1 - \mumal \\
    \Pr(\Emismatch \mid \Hhon) = \Pr(\Ematch \mid \Hmal) = \mumal
    \end{split}
\end{equation}

By performing this process, the querier can determine whether the server is honest or malicious (i.e., whether the key is valid or not), while ensuring that both false positive and false negative rates remain below predefined thresholds.

The updated cross-validation \emph{query} function, which incorporates SPRT, is presented as \proc{CrossValQuery} in Algorithm~\ref{alg:crossval-with-sprt} in Appendix~\ref{app:crossval-pseudocode}. The complete cross-validation procedure for a \emph{single target} is detailed in Algorithm~\ref{alg:cross-validation-full}, which is in Appendix~\ref{app:crossval-pseudocode}.
\begin{figure*}[htbp]
    \centering
    \includegraphics[width=0.8\linewidth]{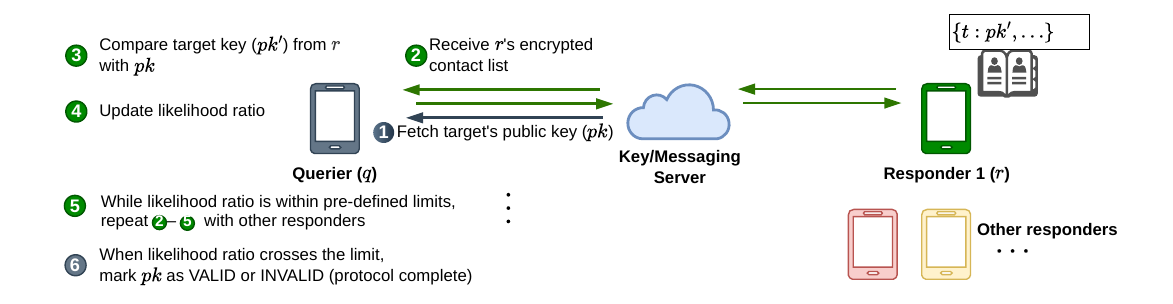}
    \caption{Key validation process in \sys. The target's key received by the querier from the server ($\mathit{pk}$) is compared with the key received from Responder 1 ($\mathit{pk}'$).}
    \label{fig:key-validation-process}
\end{figure*}

Fig.~\ref{fig:key-validation-process} illustrates the key validation process in \sys. One limitation of \sys is that when a newly joined user adds their first contact, the cross-validation process cannot be performed since no mutual contacts exist to query. In this case, the user should fall back to traditional methods such as OOB verification or KT to verify the key of the first contact.

\subsubsection{$\alpha$ and $\beta$ Selection} \label{subsubsec:alpha-beta-selection}

Setting $\alpha$ and $\beta$ to small values reduces error rates but increases the protocol failure rate of \sys. This occurs because tighter error bounds require more SPRT evidences per decision, causing validation to fail when the querier exhausts available responders to query.
Therefore, one should set $\alpha$ and $\beta$ values to reasonable levels that balance error rates with the protocol success/failure rate of \sys.

$\alpha$ not only acts as a threshold for the false positive rate but also determines the frequency of required OOB and KT checks.
Whenever \sys raises an alarm that the server may be malicious, \textbf{the user should double-check the authenticity of the server-provided key via OOB or KT mechanisms.}
Using \sys alongside OOB and KT reduces the number of required OOB and KT checks to approximately $\alpha$ (more precisely, ``$\text{(protocol failure rate)} + \text{(protocol success rate)} \times \alpha$'') when the server is honest.
According to our assumption that the server behaves honestly in most cases (due to the ``reputation'' assumption in \S\ref{subsubsec:threat-model-server}), this significantly reduces the overall frequency of OOB and KT checks.
However, as discussed earlier, setting $\alpha$ to an extremely small value may lead to high protocol failure rates, which may increase the number of required OOB and KT checks.

%% file: algorithms/crossval-nosprt.tex
\begin{algorithm}[htbp]
  \caption{Cross-Validation (No SPRT)}
  \label{alg:cross-validation-no-sprt}
  \begin{algorithmic}[1]
    \Require Querier $q$, responder $r$.
    \Require Querier's contact list $\mathcal{C}_q=\{(u, \widehat {\mathit{pk}}_u^q, s_u)\}$ where $u$ denotes a user identifier (username),
            $\widehat {\mathit{pk}}_u^q$ is a public key, and $s_u \in \{\mathsf{UNKNOWN},\mathsf{VALID},\mathsf{INVALID}\}$.
    \Require Responder's contact list $\mathcal{C}_r=\{(u, \widehat {\mathit{pk}}_u^q, s_u)\}$.
    \Require Fixed tag $\tau \in \{0,1\}^{\ell}$ used as a domain separator (payload prefix).
    \Require Primitives: \textsc{OPRF} and \textsc{OKVS}.
    \Require Notation: $\mathrm{Users}(\mathcal{C}) \triangleq \{\,u : (u,*,*)\in \mathcal{C}\,\}$.
    \medskip
    \Function{CrossValQueryNaive}{$r, \mathcal{T}$} \Comment{Executed by querier; $\mathcal{T}$ is the target subset of usernames, $r$ is the responder}
      \State \textbf{Precondition:} $\mathcal{T} \neq \emptyset \;\wedge\; \mathcal{T} \subseteq \mathrm{Users}(\mathcal{C}_q)$.
      \State \textbf{Ensure:} set of tuples $\{\, (u, \widehat {\mathit{pk}}_u^q, s_u) : u \in \mathcal{T} \,\}$ where $s_u$ is updated key status
      \State $\mathbf{y} \gets \Call{OPRF.Receive}{r,\; \mathcal{T}}$ \Comment{$\mathbf{y}$ is a map $u \mapsto y_u$} \label{line:query-oprfreceive}
      \State $\mathbf{S}_r \gets \Call{Receive}{r}$ \Comment{Receive OKVS object from responder}
      \ForAll{$u \in \mathcal{T}$}
        \State $v \gets \Call{OKVS.Decode}{\mathbf{S}_r,\; \mathbf{y}[u]}$ \label{line:query-okvsdecode}
        \If{$v = \tau \parallel \mathit{pk}'$} \Comment{Valid format: tag prefix present} \label{line:check-prefix}
          \State $s_u \gets \mathsf{VALID}$ \textbf{ if } $\mathit{pk}' = \widehat {\mathit{pk}}_u^q$ \textbf{ else } $\mathsf{INVALID}$ \label{line:check-key-match}
        \Else
          \State $s_u \gets \mathsf{UNKNOWN}$ \Comment{User $u$ not in responder's contact list}
        \EndIf
      \EndFor
      \State \Return $\{\, (u, \widehat {\mathit{pk}}_u^q, s_u) : u \in \mathcal{T} \,\}$
    \EndFunction

    \medskip
    \Function{CrossValRespond}{$q$} \Comment{Executed by responder; $q$ is the querier}
      \State $k \gets \Call{OPRF.Send}{q}$ \label{line:respond-oprfsend}
      \State $\mathbf{y} \gets \Call{OPRF.Eval}{k,\; \mathrm{Users}(\mathcal{C}_r)}$ \Comment{$u \mapsto y_u$} \label{line:respond-oprfeval}
      \State $\mathbf{M} \gets \{\, \mathbf{y}[u] \mapsto (\tau \parallel \widehat {\mathit{pk}}_u^q) \;:\; (u, \widehat {\mathit{pk}}_u^q, *) \in \mathcal{C}_r \,\}$ \label{line:respond-prefixadd}
      \State $\mathbf{S} \gets \Call{OKVS.Encode}{\mathbf{M}}$ \label{line:respond-okvsencode}
      \State \Call{Send}{$q,\; \mathbf{S}$} \label{line:respond-send}
    \EndFunction
  \end{algorithmic}
\end{algorithm}

%% file: sections/5-security-analysis.tex
\section{Security Analysis} \label{sec:security-analysis}

This section analyzes various security aspects of \sys, including the sufficiency of its probabilistic guarantees, resistance to Sybil attacks, handling of false positives and negatives, implications of collusion, the realism of the $\mumal$ assumption, prevention of brute-force attacks on OKVS, and other potential security concerns.

\subsection{Is \texorpdfstring{$1-\beta$}{1 - β} Sufficient?} \label{subsec:detection-probability}
Since \sys employs a probabilistic approach, it cannot guarantee 100\% detection of server misbehaviors. The probability of detecting a cheating server is $1 - \beta$, where $\beta$ is the user-defined false negative rate. We claim that \textbf{setting $\beta$ to an intuitively small value, such as $1\%$ or $0.5\%$, is sufficient to deter the server from attempting to cheat.} This deterrent effect relies on servers' strong incentives to maintain their reputation and avoid detection, as outlined in our threat model in \S\ref{subsubsec:threat-model-server}. If the server cheats, it will be detected with a high probability of $1 - \beta$, making the reputational risk of cheating prohibitively high for rational servers. Consequently, servers are deterred from attempting to cheat in the first place. Exceptions to this deterrent model are discussed in \S\ref{subsec:mumal-assumption}.

\subsection{Handling False Positives and Negatives} \label{subsec:false-positives-negatives}

This section details how false positives and negatives are handled in \sys. 
\paragraph{False Positives}

A false positive occurs when the querier concludes that the key server is misbehaving even if the server is honest. This can happen when the querier receives incorrect keys from a sufficient number of malicious responders. 
However, before accusing the server of its misbehavior, the querier should perform OOB verification or use KT mechanisms to confirm the authenticity of the key and whether the server is truly misbehaving. 
If the key is verified to be correct, the querier can conclude that a false positive occurred and cancel the initial conclusion. Thus, false positives do not lead to incorrect accusations against an honest server.

\paragraph{False Negatives}
A false negative occurs when the server misbehaves but the querier fails to detect it. This scenario typically arises when the server colludes with a sufficient number of malicious responders to consistently provide the same incorrect key, thus suppressing any alarm from the querier. Unfortunately, queriers cannot identify when false negatives occur. As discussed in \S\ref{subsec:detection-probability}, the messaging service providers (or their key servers) have strong reputational incentives that make them unlikely to risk attacks with a $1 - \beta$ detection probability under normal conditions. Nevertheless, if a server \emph{does} launch such an attack and successfully deceives the querier, the deception will remain undetected.

\subsection{Sybil Resistance} \label{subsec:sybil-resistance}

\sys demonstrates resistance to Sybil attacks because users typically add only known contacts (i.e., someone who they already ``know'') to their contact lists. Even if a malicious responder creates numerous accounts, it is unlikely that any single ``normal'' user will add any of these fabricated contacts. Therefore, while the global proportion of malicious users may increase, the fraction of malicious users within an individual user's contact list is likely to remain bounded.

However, a collusion remains a possibility. 
Multiple malicious contacts---potentially including the server---could collaborate to consistently return the same incorrect key to the querier. The effects of such collusions are examined in \S\ref{subsec:collusion}.

\subsection{Collusion} \label{subsec:collusion}

In this section, we analyze collusion scenarios where the \textbf{proportion of colluding users does not exceed $\mumal$}. (We will discuss cases where the proportion exceeds $\mumal$ in \S\ref{subsec:mumal-assumption}.) As discussed in \S\ref{subsec:sybil-resistance}, \sys is Sybil-resistant, and thus we assume that each service account represents a unique user. Here, we analyze two cases:

\paragraph{Case 1: $\mumal$ responders colluding with each other but not with the server}
Responders might engage in this behavior if they want the querier to \emph{falsely} accuse the server of misbehaviors. These responders can deceive the querier into believing that their provided keys are correct while the server-provided key is incorrect. Consequently, the querier may raise a false alarm (false positive). 
As discussed in \S\ref{subsec:false-positives-negatives}, the querier should verify through OOB verification or KT before accusing the server of misbehaviors, preventing false accusations against the server. In other words, the server can be protected from false accusations.

\paragraph{Case 2: $\mumal$ responders colluding with the server}
When a group of responders colludes with the server, their objective is to convince the querier that the server-provided key is genuine; thus, the responders would provide the same incorrect key as the server. In this scenario, the querier may incorrectly conclude that the server's key is authentic and fail to detect the server's misbehavior (false negative). 
As discussed in \S\ref{subsec:false-positives-negatives}, queriers cannot detect false negatives.
However, assuming that the number of malicious (colluding) responders is bounded by $\mumal$, the probability of detecting server misbehaviors remains at least $1 - \beta$, which practically deters the server from attempting such attacks, as discussed in \S\ref{subsec:detection-probability}.

\subsection{Is the \texorpdfstring{$\mumal$}{μ_mal} Assumption Realistic?} \label{subsec:mumal-assumption}

\sys's security guarantees depend critically on the assumption that the proportion of malicious users remains below $\mumal$.
As individual responders have minimal incentives to provide incorrect keys without coordination, this assumption deems reasonable in practice.

However, if a large number of users do coordinate with the server, the actual proportion of malicious responders can exceed $\mumal$ for a given querier. 
In such cases, the server may be confident that it controls sufficient responders to avoid detection; consequently, the server may launch an attack and successfully deceive the querier without being detected.
\sys provides no mechanism to detect when the ``$\mumal$ assumption'' is violated, and queriers cannot detect false negatives.
This represents a fundamental limitation: \sys's probabilistic approach works properly only when the threat model's assumptions hold; however it offers no protection when the ratio of adversaries exceeds the $\mumal$ threshold due to collusion.

\subsection{Brute-Force Attacks on OKVS} \label{subsec:okvs-brute-force}
OKVS is an \emph{offline} encoding/decoding mechanism rather than an interactive protocol. If usernames are directly used as labels, a curious client can enumerate the entire contact list by exhaustively querying every possible username. This attack becomes practical when the namespace is not large---for instance, in \signal where usernames are phone numbers.

\sys prevents this attack by using OPRF evaluations of usernames as OKVS labels instead of plaintext usernames.
Assuming the OPRF scheme is secure, it becomes computationally infeasible for the querier to guess valid OPRF outputs without querying the OPRF protocol. This means the querier \emph{must} make OPRF queries to obtain candidate labels for OKVS decoding.
The responder can impose rate limits on OPRF queries, thereby bounding the number of labels the querier can obtain for OKVS decoding.

Although OPRF is ``oblivious'' and conceals the querier's input, rate limiting remains feasible because the responder observes the querier's input set size and execution frequency. Therefore, the responder can impose rate limits on OPRF queries (e.g., a fixed quota per time window). These limits bound the number of labels the querier can obtain, effectively preventing large-scale brute-force decoding of the OKVS.

\subsection{Other Security Concerns}

\sys uses a fixed byte sequence $\tau$ as a prefix to distinguish valid public keys from random values. However, if $\tau$ is too short, it may appear by chance as the prefix of a random value (i.e., output for a non-existent label). In such cases, the querier may obtain a mismatching evidence $\Emismatch$, potentially leading to an incorrect conclusion.
If $\tau$ is $n$ bits long, the probability of such a false prefix match is $2^{-n}$. A longer prefix reduces this probability but increases bandwidth costs, creating a trade-off between security and communication efficiency. In our proof-of-concept implementation (see \S\ref{subsubsec:poc-implementation}), we set $n=56$ (i.e., 7 bytes).

%% file: sections/6-evaluation.tex
\section{Evaluation} \label{sec:evaluation}
\subsection{Simulation}
We first conduct simulations of \sys to evaluate its effectiveness using real-world datasets.
\subsubsection{Settings}
We use three datasets for simulations: \fbds, \pkds~\cite{snapnets}, and \vkds. Each dataset was collected from \facebook, \pokec, and \vk social networks, respectively.
We interpret each dataset as a \emph{directed graph} $G = (\mathcal{V}, \mathcal{E})$. In \fbds and \vkds, the friend relationship is mutual, so we treat each undirected edge as two directed edges.
Appendix~\ref{app:dataset-analysis} provides the details and analysis of each dataset.

The simulation takes $\alpha$, $\beta$, $\mumal$, and $\sigmal$ as input parameters. Here, $\sigmal$ is the probability that the key server provides a false public key upon receiving a key request. We set $\alpha = 10^{-3}$, $\beta = 10^{-2}$, $\mu_{\mathsf{mal}} = 0.05$, and $\sigmal = 0.01$. We additionally run experiments with $\mumal=0.1$ and $\mumal=0.2$ for the \vkds dataset to observe the effects of different $\mumal$ values.

Further details about the simulation settings and rationale for parameter selections are included in Appendix~\ref{app:simulation-settings}. Simulation results with different $\alpha$ and $\beta$ values are also included in Appendix~\ref{app:alpha-beta-effects}.

\subsubsection{Numbers of SPRT Iterations \& Queries} \label{subsubsec:sprt-iterations-and-queries}

\input{tables/eval-simulation.tex}

We run each simulation 5 times and record the average number of SPRT iterations ($\nsprt$) as well as the average numbers of queries ($n_{q}^{\text{unbatched}}$ and $n_{q}^{\text{batched}}$) required for the protocol to succeed.
The results when $\mumal=0.05$ are shown in Table~\ref{tab:sprt-iterations-and-queries} in the \fbds, \pkds, and \vkds columns. (In this and the subsequent section, we analyze only when $\mumal=0.05$. Thus, ignore the ``\vkds (0.1)'' and ``\vkds (0.2)'' columns for now.)
In the simulation, we consider two methods of counting queries per edge: \emph{batched query count} ($n_{q}^{\text{batched}}$) and \emph{unbatched query count} ($n_{q}^{\text{unbatched}}$). In the batched query count, each query (from a querier to a responder) is averaged by the number of targets in the batch.
Specifically, for a batched cross-validation query with target set $\mathcal{T}$, the count for each edge $(q, t)$ in the batch is incremented by $1 / |\mathcal{T}|$. In contrast, in the unbatched query count, the count for each edge is incremented by 1.

We observe that when the server is honest, the average number of required SPRT evidences (equivalent to $\nsprt$) is approximately 2.2. In contrast, when the server is malicious, the average number of required evidences is approximately 3.3. This disparity in evidence requirements stems from the imbalance between $\alpha$ and $\beta$.

The unbatched query counts in \pkds and \vkds are significantly larger than those of \fbds. This occurs because \pkds and \vkds include a substantial number of \emph{private users}. The simulation includes private users as responders, for whom we lack complete friend lists.\footnote{Private users do not publicize their contact lists, and we can only obtain partial friend lists of them via their public friends; which has a negative impact on the performance evaluation.} Private users account for 33.8\% in \pkds and 53.9\% in \vkds; see Appendix~\ref{app:dataset-analysis}. The batched query counts are significantly smaller than the corresponding unbatched query counts ($7$--$60\times$), confirming that performing OPRF queries and OKVS evaluations in batches is crucial for efficiency. Query counts for \pkds and \vkds would improve if we had complete friend lists of all private users, which constitute a significant portion of the datasets.

\subsubsection{Error/Failure Rates} \label{subsubsec:eval-error-failure-rates}

We show the average error and (protocol) failure rates in Table~\ref{tab:simulation-error-failure-rates}, in columns \fbds, \pkds, and \vkds.
An \emph{error} refers to a case where the protocol completes successfully, but the outcome does not match the ground truth (i.e., the querier reaches an incorrect conclusion about the validity of the server-provided key). This includes false positive and false negative errors. A \emph{false positive} occurs when the server is honest, but the querier's validation incorrectly identifies it as malicious. Conversely, a \emph{false negative} occurs when the server is malicious, but the validation mistakenly concludes it is honest. ``Error rates'' in the section title refers to false positive and negative rates.
A \emph{(protocol) failure} occurs when the protocol cannot reach a conclusion due to insufficient evidence (i.e., the querier has too few contacts).

Across all runs, the observed false positive and negative rates (respectively $\hat\alpha, \hat\beta$) are below the predefined thresholds ($\alpha=10^{-3}, \beta=10^{-2}$).
Protocol failure rates when the server is honest are lower than failure rates when the server is malicious across all datasets. This difference is predictable: when the server is honest, the number of required evidences ($\nsprt$) is smaller, resulting in lower failure rates. Conversely, when the server is malicious, $\nsprt$ increases, raising the likelihood of failures.

We observe protocol success rates exceeding 98.1\% when the server is \emph{honest} for both \fbds and \vkds. However, for \pkds, the protocol success rate is 52.3\%, which is significantly lower. We attribute this to the weak-tie nature of \pkds, as demonstrated in Appendix~\ref{app:dataset-analysis}.
The analysis shows that the ``number of cross-validating friends'' in \pkds is notably low with its average being only 3. Given these dataset characteristics, the low protocol success rate is not surprising.

An important consideration is that \pkds and \vkds include a substantial number of private users who do not share their friend lists. If complete friend lists of private users were available, the protocol success rates for simulations on \pkds and \vkds would improve significantly.

\fbds and \vkds achieve protocol success rates of approximately 97.2\% when the server acts \emph{maliciously}. Given the server's incentive to maintain its reputation (see \S\ref{subsec:threat-model}), such protocol success rates are acceptable. Since a cheating server is detected with probability $97.2\% \times (1 - \hat{\beta}) \approx 96\%$, \sys can effectively deter server misbehaviors.

\subsubsection{Effect of $\mumal$}

The selection of the default value $\mumal=0.05$ is two-fold. First, as \sys is Sybil-resistant (see \S\ref{subsec:sybil-resistance}), the ratio of malicious users is not likely high. Next, each responder has virtually \emph{no incentive} to attack the querier (without collusion). That is, without collusion, the proportion of malicious responders should remain sufficiently small. Therefore, we can think of situations where $\mumal$ is high as \emph{collusion} situations.

In this section, we assume situations where $\mumal$ is relatively high due to the collusion between responders and the server.
We run simulations for $\mu_{\mathsf{mal}} = 0.1$ and $\mu_{\mathsf{mal}}=0.2$ on \vkds.\footnote{Note that the simulation code already tests the worst case where all $\mumal$ users and the server collude; no additional code modifications are needed.}
Columns ``\vkds (0.1)'' and ``\vkds (0.2)'' in Tables~\ref{tab:sprt-iterations-and-queries} and \ref{tab:simulation-error-failure-rates} show the simulation results.
Obviously, as $\mumal$ increases, $\nsprt$ also increases. False positive and negative rates remain below the predefined thresholds for the $\mumal$ values.
Meanwhile, protocol failure rates increase as $\mumal$ values grow.
This demonstrates that elevated $\mumal$ leads to \emph{reduced usability of \sys} as the number of required evidences increases.

\subsection{Cost Evaluation}

\subsubsection{Proof-of-Concept (PoC) Code} \label{subsubsec:poc-implementation}

Next, we evaluate the runtime cost of \sys.
To this end, we implement \sys in Python and C++, integrated with the \textit{Matrix protocol}~\cite{thematrix.orgfoundationMatrixorg2025}\footnote{Although the source code for the \signal client/server is publicly available, hosting a \signal server on-premise is complex; thus we use \textit{Matrix} instead. \textit{Matrix} employs the \textit{Olm} cryptographic protocol for E2EE messaging~\cite{vodozemac}, which is based on the triple Diffie-Hellman (3DH) handshake and the Double Ratchet algorithm~\cite{thematrix.orgfoundationDocsOlmmdMaster2019}, similar to the \textit{\signalprotocol}.}.
To emulate a centralized messaging service, we disable \textit{Matrix}'s federation feature.
We use the \texttt{matrix-nio}~\cite{matrix-nio} Python SDK to implement clients. Messages smaller than 40~KB are sent as standard \emph{Matrix} messages. If a message exceeds 40~KB, it is sent as an encrypted attachment due to \emph{Matrix}'s event size limit.

We use the OPRF and OKVS implementations in \texttt{volePSI} library~\cite{volepsi-code}, which implements~\cite{RS21} and~\cite{RR22}.
We enable the malicious security option for OPRF to align with our threat model in \S\ref{subsec:threat-model}. The length of the prefix $\tau$ is set to 7 bytes.

To populate contact lists, we use randomly generated phone numbers as usernames and assign randomly generated raw X25519 public keys.
For simplicity, each user is assumed to have a single public key, and their username is a phone number.\footnote{In \emph{Matrix}, a username follows the format \texttt{@user\_id:homeserver.url}, and each user maintains multiple Curve25519 identity keys (one per device).}
The implementation artifacts will be made available in a future version.

\subsubsection{Experiment Settings}

We host the \emph{Matrix} server using the \emph{Synapse}~\cite{Synapse} Docker image provided by \emph{Element Inc.}~\cite{Element}, deployed on an AWS \texttt{t3.xlarge} VM instance. The instance features an \texttt{x86_64} architecture, 4 vCPUs, and 16~GiB of memory. The server is located in Stockholm, Sweden.

The PoC client code runs on two separate VMs to simulate two communicating peers. To emulate mobile-like environments, we use AWS \texttt{t4g.large} instances, which are based on the \texttt{ARM64} architecture and provide 2 vCPUs and 8~GiB of memory. The client instances are located in Seoul, South Korea.

\subsubsection{Delay}

We first measure the delay of a \emph{single \sys cross-validation query} and analyze its computational overhead. We define this measurement as the elapsed time from when the querier sends the initial message to a responder until the querier completes the comparison between the decoded value and the server-provided public key.

\begin{figure*}[htbp]
   \centering
   \begin{subfigure}{0.39\linewidth}
       \centering
       \includegraphics[width=0.8\linewidth]{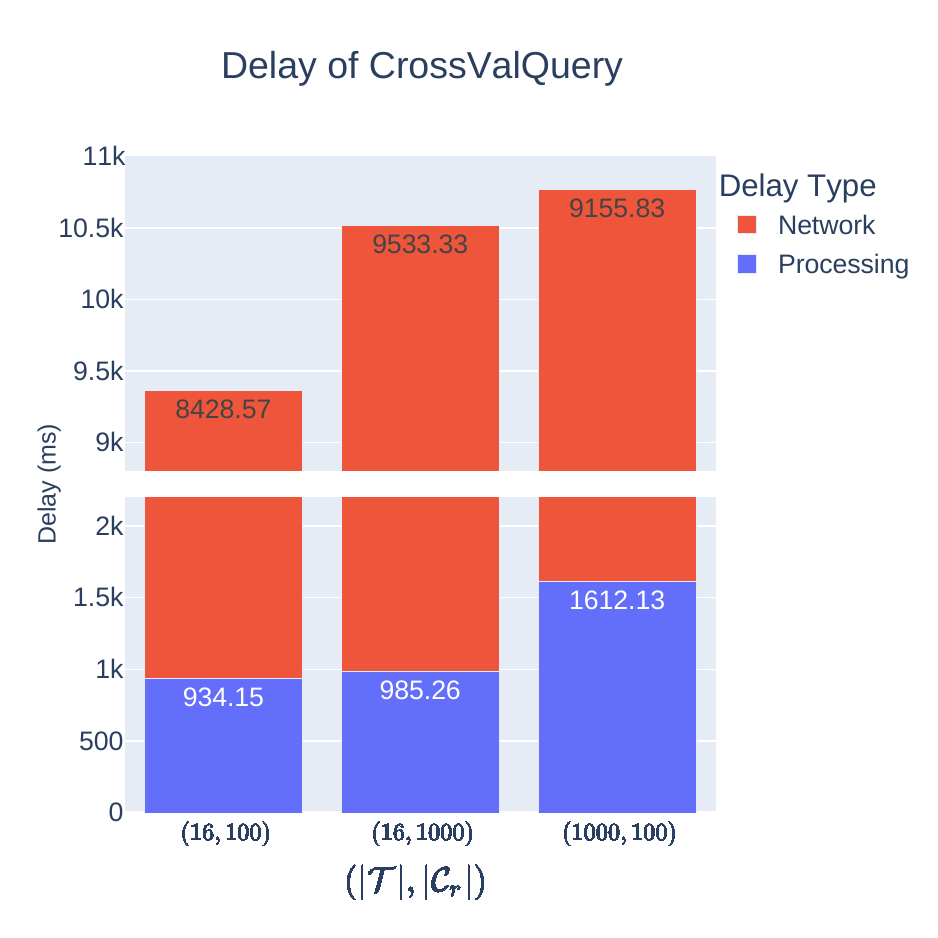}
       \caption{Completion time of \proc{CrossValQuery} for different $(|\mathcal T|,|\mathcal C_r|)$ pairs.}
       \label{fig:eval-delay-breakdown}
   \end{subfigure}
   \hfill
   \begin{subfigure}{0.6\linewidth}
       \centering
       \includegraphics[width=\linewidth]{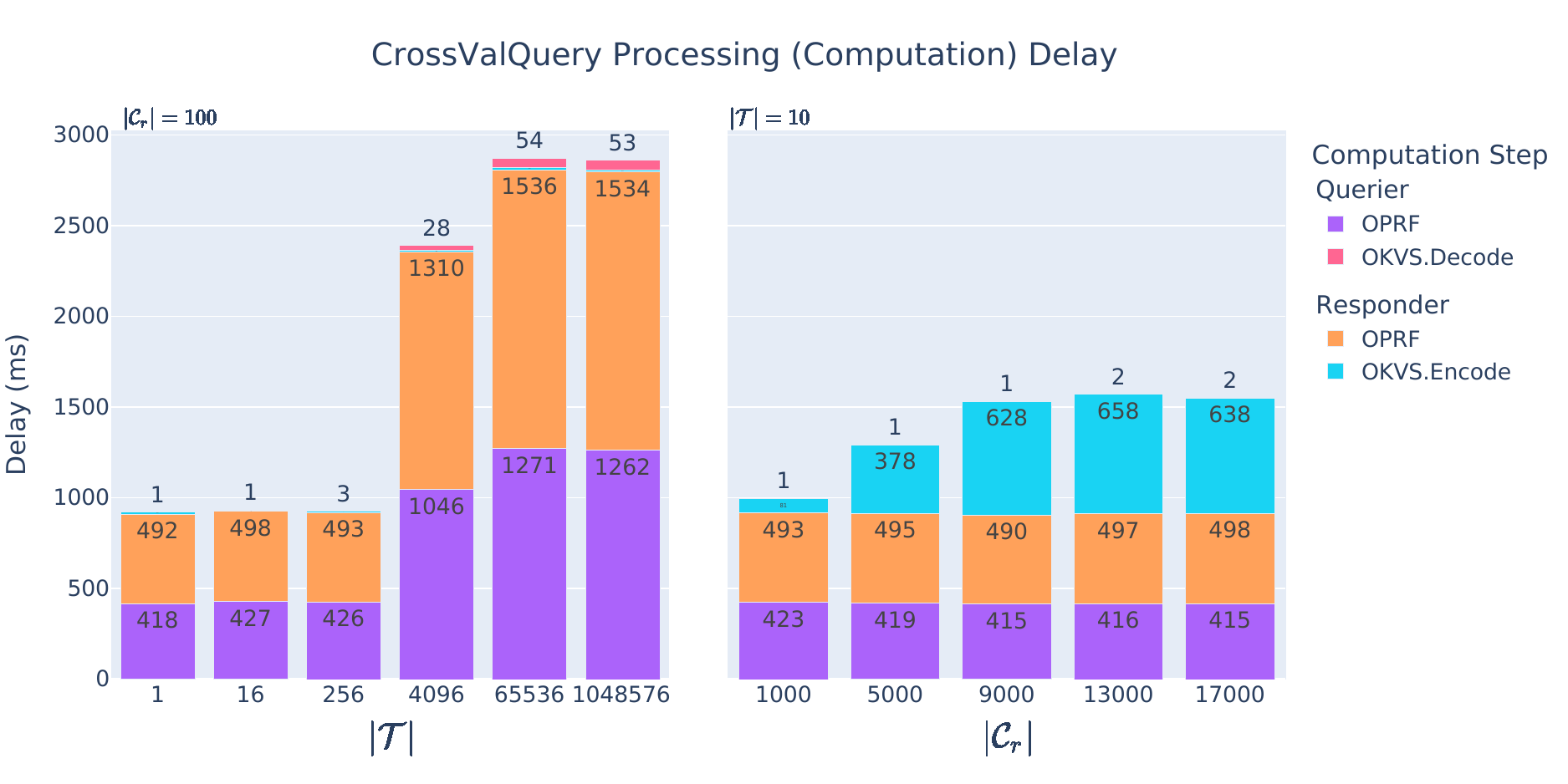}
       \caption{Processing delay of \sys \proc{CrossValQuery} operation as $|\mathcal{T}|$ or $|\mathcal{C}_r|$ varies.}
       \label{fig:crossval-processing-delay}
   \end{subfigure}
   \caption{Delay measurements for \proc{CrossValQuery} operation in \sys.}
\end{figure*}

Fig.~\ref{fig:eval-delay-breakdown} shows the evaluation results. The total delay of \proc{CrossValQuery} is 9.36~seconds when $|\mathcal{T}| = 16$ and $|\mathcal{C}_r| = 100$. We observe that network delay accounts for more than 90\% of the total delay. The extended network latency is primarily due to requiring 9--10 client-to-server round-trip times (RTTs)\footnote{It would take 4--5 RTTs if the querier and responder communicated directly, but requires twice as many client-to-server RTTs because they communicate through an intermediary server.}, for OPRF protocol execution.
Network latency also includes high transmission delay due to the large payload size (see \S\ref{subsubsec:bandwidth-cost}) and message encryption/decryption time.
Here, we include only \sys-related tasks---mainly OPRF and OKVS processing---in computational delay, while \textit{Matrix}-specific operations are included in the network delay.

To measure the delay of querying multiple responders, we conducted an additional experiment with parallel execution of \emph{five} queries (to five responders). It takes 22.86~seconds to complete five parallel queries.
Accordingly, the expected delay to validate a single key is given by $22.86 \times \lceil n_q^{\mathrm{unbatched}} / 5 \rceil$~seconds, where $n_q^{\mathrm{unbatched}}$ is the number of required queries for successful cross-validation (i.e., protocol execution) of a single key.
We calculated the average value of $22.86 \times \lceil n_q^{\mathrm{unbatched}} / 5 \rceil$ for the \fbds\footnote{We choose \fbds over \pkds and \vkds because \pkds and \vkds include a substantial number of private users with undisclosed friend lists, which significantly inflate the query counts.} dataset, which resulted in 38.55~seconds. This means that if we can run five queries in parallel, the average time required to validate a key under this setup is 38.55~seconds.

We believe the delays of \sys operations are acceptable for background operations, as users can continue using their messaging application while key validation completes in the background.

Fig.~\ref{fig:crossval-processing-delay} shows the \emph{processing} delay of \proc{CrossValQuery}, broken down by computational tasks.
In all scenarios, OPRF computations are the most time-consuming step for both the querier and the responder. When $|\mathcal{T}|$ and $|\mathcal{C}_r|$ are small, OPRF computations dominate the overall processing delay.
As $|\mathcal{T}|$ increases, the cost of OPRF computations also increases but converges. This occurs because the OPRF protocol used in the PoC is based on a one-time vector OLE (VOLE) construction, which is optimized for large input sizes. Since the typical OPRF input size in \sys is relatively small for messaging applications, another OPRF protocol optimized for small inputs would be more suitable.
Similarly, as $|\mathcal{C}_r|$ grows, the cost of OKVS encoding by the responder increases.

\subsubsection{Bandwidth Cost} \label{subsubsec:bandwidth-cost}

We then measure the bandwidth consumption of \sys by measuring the total size of the application-layer payload.

\paragraph{Cost for a single query}
We obtain the following bandwidth cost for a single \proc{CrossValQuery} execution (see Appendix~\ref{app:bandwidth-eval-details} for details):
\begin{itemize}
    \item Querier-to-responder: 550~KB (constant)
    \item Responder-to-querier: $0.0521 \times |\mathcal{C}_r| + 81.76$~KB
\end{itemize}

We observe that the querier-to-responder traffic is notably higher than the responder-to-querier traffic. The querier, who benefits from the operation, bears higher network costs by sending more data than received compared to the responder. Given that most home networks and mobile devices have greater download bandwidth capacity than upload capacity, this asymmetry allows responders to participate in \sys with reduced overhead.

\paragraph{Cost for validating a key}
Next, we calculate the bandwidth cost of validating a single key by making multiple cross-validation queries. For this calculation, we assume that $|\mathcal{C}_r| = 100$ and $|n_{q}^{\text{unbatched}}|=6.83$, based on the simulation results for \fbds in \S\ref{subsubsec:sprt-iterations-and-queries}. The average bandwidth consumption for each \emph{querier} is:
\begin{itemize}
    \item Upload: $6.83 \times 550$~KB = 3.76~MB
    \item Download: $6.83 \times (0.0521 \times 100 + 81.76)$~KB = 594~KB
\end{itemize}

Each \emph{responder}, on average, generates traffic:
\begin{itemize}
    \item Upload: $0.0521 \times 100 + 81.76 = 86.97$~KB
    \item Download: 550~KB
\end{itemize}

We claim that \sys's bandwidth cost is acceptable for modern mobile networks and falls well within practical limits.

%% file: tables/eval-simulation.tex
\begin{table*}[htbp]
    \centering
    \caption{Simulation results of \sys on three datasets (\fbds, \pkds, \vkds) with $\alpha=10^{-3}$, $\beta=10^{-2}$, $\sigmal=0.01$. Configuration columns indicate dataset and $\mumal$ value in format ``\texttt{dataset} ($\mumal$)''; empty $\mumal$ field indicates $\mumal = 0.05$.}
    \begin{subtable}{0.7\linewidth}
        \centering
        \caption{Average number of SPRT iterations ($\nsprt$) and queries required for successful protocol completion. Query counts use batched ($n_{q}^{\mathrm{batched}}$) and unbatched ($n_{q}^{\mathrm{unbatched}}$) methods.}
        \label{tab:sprt-iterations-and-queries}
        \begin{tabular}{l|l|rrrrr}
            \toprule
            \multirow{2}{*}{\makecell[l]{\textbf{Server status}}} & \multirow{2}{*}{\textbf{Metric}} & \multicolumn{5}{c}{\textbf{Configuration}}                                              \\
            \cline{3-7}
                                                                  &                                  & \fbds                                      & \pkds & \vkds  & \vkds (0.1) & \vkds (0.2) \\
            \midrule
            \multirow{3}{*}{\makecell[l]{Honest                                                                                                                                                \\
            ($\widehat{\mathit{pk}_{(q,t)}}=pk_t$)}}                       & $\nsprt$                         & 2.20                                       & 2.13  & 2.22   & 3.74        & 6.55        \\
                                                                  & $n_{q}^{\text{unbatched}}$       & 6.83                                       & 33.48 & 81.23  & 118.84      & 173.47      \\
                                                                  & $n_{q}^{\text{batched}}$         & 0.41                                       & 0.56  & 8.65   & 9.22        & 9.41        \\
            \hline
            \multirow{3}{*}{\makecell[l]{Malicious                                                                                                                                             \\
            ($\widehat{\mathit{pk}_{(q,t)}} = pk_t^{\mathsf{mal}}$)}}      & $\nsprt$                         & 3.28                                       & 3.19  & 3.31   & 4.96        & 8.12        \\
                                                                  & $n_{q}^{\text{unbatched}}$       & 9.36                                       & 41.26 & 111.37 & 150.73      & 194.64      \\
                                                                  & $n_{q}^{\text{batched}}$         & 0.78                                       & 0.74  & 14.83  & 12.56       & 11.03       \\
            \bottomrule
        \end{tabular}
    \end{subtable}
    \hfill
    \begin{subtable}{0.7\linewidth}
        \centering
        \caption{Average error and (protocol) failure rates across simulations. Observed error rates remain below expected thresholds ($\alpha=0.001, \beta=0.01$) in all cases.}
        \label{tab:simulation-error-failure-rates}
        \begin{tabular}{lrrrrr}
            \toprule
            \multirow{2}{*}{\textbf{Metric}}     & \multicolumn{5}{c}{\textbf{Configuration}}                                               \\
            \cline{2-6}
                                                 & \fbds                                      & \pkds  & \vkds  & \vkds (0.1) & \vkds (0.2) \\ \midrule
            False positive rate ($\hat{\alpha}$) & 0.0002                                     & 0.0001 & 0.0001 & 0.0002      & 0.0009      \\
            False negative rate ($\hat{\beta}$)  & 0.0034                                     & 0.0014 & 0.0029 & 0.0012      & 0.0035      \\
            \hline
            Failure rate (server honest)         & 0.0133                                     & 0.5227 & 0.0184 & 0.0327      & 0.0579      \\
            Failure rate (server malicious)      & 0.0268                                     & 0.6295 & 0.0289 & 0.0449      & 0.0712      \\
            \bottomrule
        \end{tabular}
    \end{subtable}
\end{table*}

%% file: sections/7-discussion.tex
\section{Discussion} \label{sec:discussion}
\subsection{Handling Key Updates}

\sys is vulnerable to misinterpretation when the target's public key has been recently updated. For instance, if the querier retrieves the updated key from the key server but the responder still holds the target's previous key, the querier may incorrectly suspect that the server has returned a tampered key.
A straightforward mitigation is for the querier to store the target's previous keys and ignore responses containing outdated ones. Notice that this approach requires queriers to maintain storage for old key versions, or it may suffer from a ``no key'' case (i.e., the querier simply didn't have the target in their contact list in that past time point).

For this, \sys is extended to make the server store $(\mathsf{username}, \mathsf{version} \parallel \mathsf{public\_key})$ instead of $(\mathsf{username}, \mathsf{public\_key})$. Unlike KT, the server in \sys maintains only the current version of each key.
When Alice queries Bob's key, the server responds with a 3-tuple: Bob's key, its version number, and the server's signature over:
\[ \mathsf{Bob\_username} \parallel \mathsf{version} \parallel \mathsf{public\_key} \parallel \mathsf{Alice\_username}.\]
Alice verifies the signature to confirm its authenticity.
Alice stores this 3-tuple. When she later \emph{responds} to a cross-validation request from her contact as a responder, she encodes and sends this tuple as an OKVS entry (instead of just the public key).
The querier (or the contact) verifies the signature to confirm that the key was sent legitimately from the server.

Suppose the querier requests Bob's public key from both the server and Alice (the responder).
The querier then compares the version numbers of the two keys.
\begin{itemize}
   \item If the versions match, the response is accepted.
   \item If the responder's version is older, the response is ignored.
   \item If the responder's version is newer, the querier suspects the server has provided an outdated key. (It will fall back to OOB or KT.)
\end{itemize}

\subsection{Protocol Availability}

\sys is an interactive protocol  between a querier and multiple responders, raising availability concerns since the fundamental assumption of E2EE messaging services is that users may go offline. However, since \sys uses only the current user's contact list as an input (for both querier and responder) and requires no manual intervention, it can operate seamlessly in the background as part of the messaging application. Consequently, a responder can answer a querier's request as long as their device remains powered on and connected to the Internet. If a responder does not reply, the querier can simply select another responder and retry the cross-validation later if necessary. Although evaluating battery consumption on mobile devices remains future work, the availability issue is not important.

\subsection{Privacy}

While \sys employs OPRF and OKVS to preserve query privacy, several privacy concerns remain. First, the responder inevitably learns that the querier has added them to their contact list. This may be undesirable, as the querier might wish to keep their contact selections private.

Second, the querier could attempt to infer the intersection of their contact list with the responder's by submitting their entire contact list as the OPRF input set. Even if the responder enforces a limit on the OPRF input size to prevent brute-force queries, the querier can still test many targets up to the limit. To mitigate this, the responder should carefully select the maximum allowed input size and apply rate-limiting to queries.

Third, the querier may estimate the size of the responder's contact list by analyzing the encoded OKVS object, whose size grows approximately linearly with the number of encoded key-value pairs. To reduce this leakage, the responder can add dummy key-value pairs when constructing the OKVS object.
The more dummy pairs added, the more difficult it becomes for the querier to infer the actual list size. However, this also increases the OKVS object size and corresponding communication overhead.

\subsection{Limitations}

While \sys provides a practical approach to key validation through social graphs, several limitations need to be discussed:

First, \sys faces the traditional bootstrapping problem: the system cannot validate keys when users lack mutual contacts with their intended targets. This is particularly problematic for new users adding their first contact, who must rely on traditional OOB verification or KT queries.

Second, \sys has limited effectiveness on social networks dominated by \emph{weak ties}.
As demonstrated in \S\ref{subsubsec:eval-error-failure-rates}, our simulations reveal that \sys performs poorly on networks characterized by weak social ties. For example, the \pkds dataset---which includes many commercial accounts and unidirectional ``following'' relationships---achieves only a 52.3\% protocol success rate when the server is honest, compared to over 98\% for \fbds and \vkds. This suggests that \sys is unsuitable for platforms where users primarily maintain weak connections rather than strong social ties.

Third, as discussed in \S\ref{subsec:false-positives-negatives}, \sys cannot recover from false negative errors.
Although our threat model assumes that a server's concern for reputation deters attacks that would be detected with high probability $(1-\beta)$, any successful attack within the $\beta$ probability window remains permanently undetected.

Finally, as discussed in \S\ref{subsec:mumal-assumption}, \sys is vulnerable to collusion attacks that violate its core assumption about the fraction of malicious responders (i.e., $\mumal$).
Under typical scenarios this assumption holds; however, coordinated collusion attacks break this foundational premise and undermine the protocol's security guarantees.

%% file: sections/8-future-work.tex
\section{Future Work} \label{sec:future-work}

There are several promising directions for extending and improving \sys.
First, the current implementation uses an OPRF protocol from~\cite{volepsi-code} that is optimized for large batches. Future work should investigate alternative OPRF protocols specifically designed for small input sizes, which are more suitable for \sys.

Second, we plan to explore the design of alternative key directory formats. Since \sys reduces the frequency of KT queries, it enables investigating alternative KT directory formats that prioritize storage efficiency over query processing delays, although such alternatives are not explored in this work. An alternative directory could achieve orders-of-magnitude reductions in storage costs at the expense of higher query processing latency while preserving the same cryptographic security guarantees as traditional Merkle tree-based directories. Potential approaches include RSA accumulator-based designs such as~\cite{kemmoeRSABasedDynamicAccumulator2024}.

Finally, a comprehensive analysis of battery consumption should be conducted. Because \sys is designed to run continuously in the background on mobile devices, evaluating its impact on a battery life is critical for assessing its practical applicability.

%% file: sections/9-related-work.tex
\section{Related Work} \label{sec:related-work}

\subsection{Key Authentication in E2EE Messaging}

\paragraph{Out-of-Band (OOB) Verification}
\textit{Shinrvanian et al.}~\cite{shirvanianPitfallsEndtoendEncrypted2017} show that out-of-band key verification methods, such as \signal's safety numbers and \whatsapp's security codes, suffer from poor usability, especially in settings where two users are not in proximity.

\paragraph{Key Transparency (KT) Systems}
\textit{CONIKS}~\cite{CONIKS} pioneered cryptographically verifiable key directories using Merkle trees and consistency proofs. However, its storage requirements scale linearly with epochs, making deployment impractical for billion-user systems. \textit{SEEMless}~\cite{SEEMless} and \parakeet~\cite{Parakeet} reduce storage to scale with key updates rather than epochs. Recent systems optimize various aspects: \textit{OPTIKS}~\cite{OPTIKS} halves storage requirements compared to \parakeet through improved data structures, while \textit{ELEKTRA}~\cite{ELEKTRA} addresses multi-device scenarios with formal security proofs.

\subsection{Social Graph-Based Security}
\textit{PGP's web of trust (WoT)} represents an early attempt at social key validation. WoT allows users to sign each other's keys, creating a decentralized trust network.
However, the WoT has seen limited practical participation and deployment---only a small subset of keys can fully benefit from it---and the model is not user-friendly.
\textit{Ulrich et al.}~\cite{ulrichInvestigatingOpenPGPWeb2011} analyze WoT structure, verifiability, and robustness, while \textit{Barenghi et al.}~\cite{barenghiChallengingTrustworthinessPGP2015} examine security weaknesses that can undermine WoT assurances.

\emph{Trusted Introductions (TI)}~\cite{gloorTrustedIntroductionsFornbspsecure2023} has conceptual similarities to our work: it leverages social contacts for key verification. \textit{TI} proposes forwarding identity verifications across already-verified channels in \signal via a mutual friend (or the ``introducer''). However, \textit{TI} and \sys are fundamentally different in their goals; \textit{TI} proposes a usability/UX improvement to facilitate out-of-band key verification through mutual friends, while \sys is a cryptographic protocol for privacy-preserving key validation using social contacts.

\subsection{OPRF and OKVS}

\paragraph{Oblivious Pseudorandom Functions (OPRFs)}
OPRFs have been widely studied as a primitive for private set intersection (PSI), password hardening, contact discovery, and more.
\textit{Casacuberta et al.}~\cite{SoK-OPRF} present a systematization of OPRF constructions, classifying them into families and surveying their applications. 
RFC 9497 \cite{rfc9497} specifies interoperable OPRF and VOPRF (verifiable OPRF) protocols instantiated in standard prime-order groups, including elliptic curves.
\textit{Tyagi et al.}~\cite{tyagiFastSimplePartially2022} construct a partially-oblivious PRF (allowing a public input) without using expensive pairings, enabling new applications like Privacy Pass and OPAQUE with minimal overhead.

The PoC implementation of our protocol builds on the OPRF and OKVS constructions from \cite{RS21} and \cite{RR22}.
The first work, \textit{VOLE-PSI}~\cite{RS21}, leverages a batched OPRF built from vector oblivious linear evaluation (VOLE) to achieve state-of-the-art efficiency in PSI.
A follow-up work, \cite{RR22}, improves the VOLE performance for OPRF and PSI protocols in \cite{RS21}. It also enhances the computation and communication efficiency of the OKVS data structure used in \cite{RS21}.

\paragraph{Oblivious Key-Value Stores (OKVSs)}
\textit{Garimella et al.}~\cite{OKVS} first introduced the OKVS concept, showing it can significantly reduce the communication in PSI by combining hashing with secret-sharing techniques.
Follow-up studies improved OKVS efficiency and utility. Notably, \textit{Bienstock et al.}~\cite{bienstockNearoptimalObliviousKeyvalue2023a} construct a near-optimal OKVS with fast encoding/decoding, enabling state-of-the-art PSI with lower bandwidth and supporting volume-hiding encrypted multi-maps with less storage and much faster queries than prior designs, improving \cite{OKVS,RR22}.

\subsection{Alternative Authenticated Data Structures}

\sys effectively reduces the rate of KT queries by orders, enabling to use alternative KT directory formats that prioritize storage efficiency over query-processing performance. An \textit{RSA accumulator} is such an alternative.
RSA accumulators provide constant-size set representations with membership proofs, offering potential storage improvements over Merkle trees. 
\textit{Kemmoe et al.}~\cite{kemmoeRSABasedDynamicAccumulator2024} eliminate prime generation requirements, while \textit{Baldimtsi et al.}~\cite{baldimtsiObliviousAccumulators2024} introduce oblivious variants preserving privacy.

%% file: sections/10-conclusion.tex
\section{Conclusion} \label{sec:conclusion}

This paper presents \sys, a protocol that leverages social graphs to cross-validate public keys in end-to-end encrypted messaging systems. By enabling users to verify keys through mutual contacts while preserving privacy via OPRF and OKVS, \sys addresses the limitations of (i) highly trustworthy but non-scalable out-of-band key verification and (ii) scalable but storage-intensive key transparency systems.

Our evaluation demonstrates that \sys achieves detection rates exceeding 97\% against key substitution attacks in strong-to-moderate-tie social networks, while maintaining false positive rates below 0.1\% and false negative rates below 0.3\%. 
Our proof-of-concept implementation confirms that a key validation completes within acceptable background processing time and bandwidth cost on commodity hardware.

\sys's primary contribution is not replacing existing key authentication methods, but rather reducing their required frequency by approximately two orders of magnitude. This dramatic reduction enables fundamental architectural transformations in key transparency systems: directories can transition to low-cost storage solutions or adopt space-efficient data structures that trade storage for higher query processing costs.
By augmenting rather than replacing existing methods, \sys provides a practical pathway to scalable key authentication that avoids both prohibitive server-side storage requirements and excessive manual intervention.

%% file: sections/appendix.tex
\input{sections/appendix/akd-storage-cost.tex}
\input{sections/appendix/alg-notations.tex}
\input{sections/appendix/sprt.tex}
\input{sections/appendix/crossval-pseudocode.tex}
\input{sections/appendix/dataset-analysis.tex}
\input{sections/appendix/simulation-settings.tex}

\input{sections/appendix/bandwidth-cost-eval.tex}

%% file: sections/appendix/akd-storage-cost.tex
\section{\akd Storage Cost} \label{app:akd-storage-cost}

In this section, we evaluate the server-side storage cost of \akd~\cite{AKD}, a key transparency implementation based on \textit{SEEMless}~\cite{SEEMless} and \parakeet~\cite{Parakeet}. We write a small program that inserts $x$ unique key-value pairs into the \akd directory and measure the heap memory usage of the program. Each label (key) is a randomly generated phone number, and each value is a 32-byte fixed-length sequence (equal to the size of a raw X25519 public key).

\begin{figure}[htbp]
    \centering
    \includegraphics[width=0.65\linewidth]{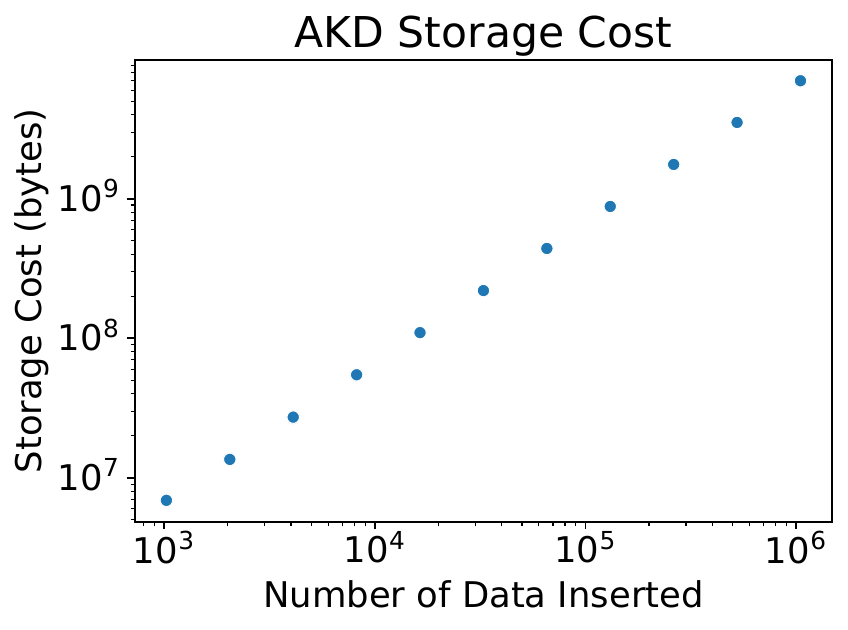}
    \caption{\akd storage cost evaluation results. Storage cost (bytes) versus number of inserted entries.}
    \label{fig:akd-storage-cost}
\end{figure}

Fig.~\ref{fig:akd-storage-cost} shows the result. As expected, the storage cost grows linearly with the number of inserted entries. 
The linear regression result is as follows ($x$ is the number of data inserted, and $y$ is the storage cost):
$$y = 6.715x - 159.9 \text{~KB}$$

\begin{table}
    \centering
    \caption{Estimated \akd storage costs for popular messaging services assuming four key versions per user. ``MAUs'' denotes monthly active users.}
    \label{tab:akd-storage-cost}
    \begin{tabular}{lrr}
        \hline
        \textbf{Service} & \textbf{MAUs} & \textbf{Storage Cost} \\
        \hline
        \whatsapp & $\sim 3$ billion \cite{mehtaWhatsAppNowHas2025} & 80.6~TB \\
        \telegram & $\sim 1$ billion \cite{mehtaTelegramFounderPavel2025} & 26.9~TB \\
        \signal & $\sim 40$--70 million \cite{tomgerkenWhatMessagingApp2025} & 1.1--1.9~TB \\
        \hline
    \end{tabular}
\end{table}

In real-world deployments, \akd must store the \emph{previous versions} of keys, and its theoretical storage cost grows linearly with the \emph{total number of key updates}. A typical user may update their key multiple times (e.g., due to device changes or application re-installs). Assuming each user has \emph{four} versions stored in the \akd directory, the practical storage cost of \akd considering the number of monthly active users (MAUs) in popular messaging services is shown in Table~\ref{tab:akd-storage-cost}. The storage cost of \akd is substantial, especially for popular messaging services like \whatsapp and \telegram, which have billions of users.

%% file: sections/appendix/alg-notations.tex
\section{Algorithm Notations} \label{app:alg-notations}

This section provides a summary of API semantics and notations used by algorithms in this paper.

\subsection{API Semantics}

\subsubsection{OPRF (Oblivious Pseudorandom Function)}
\begin{itemize}
\item \textproc{OPRF.Receive}$(r,\mathcal{T})$ : The client-side API for OPRF (see \S\ref{subsubsec:oprf}), executed by the \emph{querier}. $\mathcal{T}$ is the OPRF input set (i.e., a set of target usernames). $r$ is the responder against whom the OPRF protocol is executed.
\item \textproc{OPRF.Send}$(q)$: The server-side API for OPRF, executed by the \emph{responder}. It listens for a querier $q$ to initiate an OPRF session and executes the OPRF protocol. Consequently, it derives a key $k$ and returns $k$ to local context for use in \textsc{OPRF.Eval}; it does not reveal $k$ to the querier. 
\item \textproc{OPRF.Eval}$(k,U)$: Deterministically computes $y_u = PRF_k(u)$ for each $u \in U$ under the key $k$ previously derived by \textproc{OPRF.Send} and returns the mapping $u\mapsto y_u$; for the same $k$ and $u$, the output is identical across calls.
\end{itemize} 

Note that \textproc{OPRF.Receive} and \textproc{OPRF.Send} include transport functionality, similar to \textproc{Send} and \textproc{Receive} in \S\ref{app:alg-transport-primitives}, to send and receive messages between the querier and responder.

\subsubsection{OKVS (Oblivious Key-Value Store)}
\begin{itemize}
   \item \textproc{OKVS.Encode}$(\mathbf{M})$: Corresponds to the \textsf{Encode} function in \S\ref{subsubsec:okvs}. It takes a \emph{map} as input instead of a set of key-value pairs, in contrast to \textsf{Encode}. It outputs the encoded OKVS object $\mathbf{S}$, or $\bot$ with statistically small probability.
   \item \textproc{OKVS.Decode}$(\mathbf{S}, y)$: Corresponds to the \textsf{Decode} function in \S\ref{subsubsec:okvs}. It takes an encoded OKVS object $\mathbf{S}$ and a key $y$ as input, and returns the value $v$ associated with $y$ in $\mathbf{S}$, or a random value if $y$ is not present in $\mathbf{S}$.
\end{itemize} 

\subsubsection{General Transport Primitives} \label{app:alg-transport-primitives}
\begin{itemize}
   \item \textproc{Send}$(q,\mathbf{S})$: General transport primitive to post a payload to the peer. It is invoked by the \emph{responder} to send the encoded OKVS object $\mathbf{S}$ to the querier $q$.
   \item \textproc{Receive}$(r)$: General transport primitive to receive a payload from the peer. It is invoked by the \emph{querier} to receive the encoded OKVS object from the responder $r$.
\end{itemize}

\subsubsection{SPRT (Sequential Probability Ratio Test)}
\begin{itemize}
   \item \textproc{SPRT}$(\alpha,\beta,\mu_{\mathsf{mal}},\ \Sigma_u,\ e)$: This function implements the SPRT algorithm as described in \S\ref{subsec:crossval-sprt}. 
   \subitem \textbf{Input parameters:} $\alpha$, $\beta$, $\mumal$, the current SLLR (cumulative log-likelihood ratio) $\Sigma_u$ for a user $u$, and the observed evidence $e$ (either $\Ematch$ or $\Emismatch$). 
   \subitem \textbf{Return value:} A tuple $(\Sigma_u', \delta)$. $\Sigma_u'$ is the updated SLLR, and $\delta \in \{\mathsf{VALID}, \mathsf{INVALID}, \mathsf{UNKNOWN}\}$ is the conclusion. Note that $\mathsf{UNKNOWN}$ is returned if the SPRT test is inconclusive, indicating that the evidence is insufficient to make a decision.
\end{itemize}

%% file: sections/appendix/sprt.tex
\section{Sequential Probability Ratio Test (SPRT)} \label{app:sprt}

The \emph{sequential probability ratio test (SPRT)} is a statistical method for hypothesis testing in a sequential setting. Unlike fixed-sample tests, which require a pre-specified number of observations, SPRT evaluates data as it is collected and may reach a decision earlier, potentially reducing the number of required samples.

SPRT is used to distinguish between two simple hypotheses:

\[
    H_0: \theta = \theta_0 \quad \text{versus} \quad H_1: \theta = \theta_1,
\]

where \( \theta \) is a parameter governing the distribution of the observations. As data points \( x_1, x_2, \dots, x_n \) are observed sequentially, the \emph{likelihood ratio} is computed at each step:

\[
    \Lambda_n = \prod_{i=1}^{n} \frac{f(x_i; \theta_1)}{f(x_i; \theta_0)},
\]

or, equivalently, in logarithmic form:

\[
    \log \Lambda_n = \sum_{i=1}^{n} \log \left( \frac{f(x_i; \theta_1)}{f(x_i; \theta_0)} \right),
\]

where \( f(x_i; \theta) \) is the probability density or mass function under parameter \( \theta \).

The decision rule is defined as follows:
\begin{itemize}
    \item If \( \log \Lambda_n \ge a \), accept \( H_1 \)
    \item If \( \log \Lambda_n \le b \), accept \( H_0 \)
    \item If \( b < \log \Lambda_n < a \), continue sampling
\end{itemize}

The thresholds \( a \) and \( b \) are determined by the required error probabilities:

\[
    a = \log{\frac{1 - \beta}{\alpha}}, \quad b = \log{\frac{\beta}{1 - \alpha}},
\]

where \( \alpha \) is the maximum acceptable Type I error (false positive rate), and \( \beta \) is the maximum acceptable Type II error (false negative rate).

%% file: sections/appendix/crossval-pseudocode.tex
\section{Full Cross-Validation Pseudocode} \label{app:crossval-pseudocode}

\input{algorithms/crossval-sprt}

Algorithm~\ref{alg:crossval-with-sprt} presents the pseudocode for the \proc{CrossValQuery} function integrating SPRT. Reference for API semantics and notations can be found in Appendix~\ref{app:alg-notations}.

\input{algorithms/crossval-full}

Algorithm~\ref{alg:cross-validation-full} presents the complete cross-validation procedure triggered when a contact's key is changed or a new contact is added. To fully leverage batch querying, the querier sends queries not only for the new or updated contact but also for existing contacts whose keys remain unverified. To avoid collecting duplicate pieces of evidence, the querier maintains a record of responders that have already contributed evidence for specific targets and excludes them from subsequent queries. In \proc{OnContactAdd}, when a new contact is successfully validated, the querier invokes \proc{CrossValQuery} once for the remaining unverified targets, using the newly validated contact as a responder (Lines~\ref{line:on-contact-add-if-start}--\ref{line:on-contact-add-if-end}). This approach is based on the possibility that the new contact may have any targets in their contact list and thus may be able to provide new evidence for previously unverified targets.

%% file: algorithms/crossval-sprt.tex
\begin{algorithm}[htbp]
  \caption{Cross-Validation (with SPRT)}
  \label{alg:crossval-with-sprt}
  \begin{algorithmic}[1]
    \Require Querier $q$, responder $r$.
    \Require $\alpha,\beta, \mu_{\mathsf{mal}} \in (0, 1)$
    \Require Querier's contact list $\mathcal{C}_q=\{(u, \widehat {\mathit{pk}}_u^q, s_u, \mathrm{Prog}_u)\}$ where
            $u$ is a username, $\widehat {\mathit{pk}}_u^q$ a public key, $s_u \in \{\mathsf{UNKNOWN},\mathsf{VALID},\mathsf{INVALID}\}$,
            and $\mathrm{Prog}_u=(\Sigma_u,\ \mathrm{EvUsers}_u)$ with $\Sigma_u\in\mathbb{R}$ the running SLLR and
            $\mathrm{EvUsers}_u$ a set of responders who contributed evidence.
    \Require Fixed tag $\tau \in \{0,1\}^{\ell}$ used as a domain separator (payload prefix).
    \Require Primitives: \textsc{OPRF} and \textsc{OKVS}.
    \Require Notation: $\mathrm{Users}(\mathcal{C}) \triangleq \{\,u : (u,*,*,*)\in \mathcal{C}\,\}$.
\medskip
    \Function{CrossValQuery}{$r, \mathcal{T}$} \Comment{Executed by querier; $\mathcal{T}$ is the target subset of usernames, $r$ is the responder}
    \State \textbf{Precondition:} $\mathcal{T} \neq \emptyset \;\wedge\; \mathcal{T} \subseteq \mathrm{Users}(\mathcal{C}_q)$.
    \State \textbf{Ensure:} set of usernames $\mathcal{W} \subseteq \mathcal{T}$ that have been finalized (verified) this round
      \State $\mathbf{y} \gets \Call{OPRF.Receive}{r,\; \mathcal{T}}$ \Comment{$\mathbf{y}$ is a map $u \mapsto y_u$}
      \State $\mathbf{S}_r \gets \Call{Receive}{r}$ \Comment{OKVS object from $r$}
      \State $\mathcal{W} \gets \emptyset$ \Comment{finalized users this round}
      \ForAll{$u \in \mathcal{T}$}
        \If{$r \in \mathrm{EvUsers}_u$} \textbf{continue}
        \EndIf
        \State $v \gets \Call{OKVS.Decode}{\mathbf{S}_r,\; \mathbf{y}[u]}$ \Comment{$v \in \{0,1\}^*$}
        \If{$v = \tau \parallel \mathit{pk}'$} \Comment{tagged payload, valid entry}
          \State $e \gets \Ematch \ \textbf{ if } \ \mathit{pk}'=\widehat {\mathit{pk}}_u^q \ \textbf{ else } \ \Emismatch$
          \State $\mathrm{EvUsers}_u \gets \mathrm{EvUsers}_u \cup \{r\}$
          \State $(\Sigma_u',\ \delta) \gets \Call{SPRT}{\alpha,\beta,\mu_{\mathsf{mal}},\ \Sigma_u,\ e}$
          \State $\Sigma_u \gets \Sigma_u'$ \Comment{update SLLR}
          \If{$\delta \in \{\mathsf{VALID},\mathsf{INVALID}\}$}
            \If{$\delta=\mathsf{INVALID}$}  \textbf{raise alarm} \Comment{server may be malicious}
            \EndIf
            \State $s_u \gets \delta$
            \State $\mathcal{W} \gets \mathcal{W} \cup \{u\}$
          \EndIf
        \Else
          \State $s_u \gets \mathsf{UNKNOWN}$ \Comment{no entry for this username}
        \EndIf
      \EndFor
      \State \Return $\mathcal{W}$
    \EndFunction
\medskip

    \Function{CrossValRespond}{$q$} \Comment{Executed by responder; $q$ is the querier}
      \State Same as Algorithm~\ref{alg:cross-validation-no-sprt}
    \EndFunction
  \end{algorithmic}
\end{algorithm}

%% file: algorithms/crossval-full.tex
  \begin{algorithm}[htbp]
  \caption{Full Cross-Validation Process}
  \label{alg:cross-validation-full}
  \begin{algorithmic}[1]
    \Require Querier's contact list $\mathcal{C}_q=\{(u, \widehat {\mathit{pk}}_u^q, s_u, \mathrm{Prog}_u)\}$
    \Statex where $s_u \in \{\mathsf{UNVERIFIED},\mathsf{VALID},\mathsf{INVALID}\}$ and
    \Statex $\mathrm{Prog}_u=(\Sigma_u,\ \mathrm{EvUsers}_u)$ with $\Sigma_u\in\mathbb{R}$ (running SLLR) and $\mathrm{EvUsers}_u\subseteq\mathrm{Users}(\mathcal{C}_q)$.
    \Require Notation: $\mathrm{Users}(\mathcal{C}_q) \triangleq \{\,u : (u,*,*,*)\in \mathcal{C}_q\,\}$.
\medskip
    \Function{OnKeyChange}{$u$} \Comment{executed by the querier; $u$ is the contact whose key changed}
      \State \textbf{Precondition:} $u \in \mathrm{Users}(\mathcal{C}_q)$
      \State \textbf{Ensure:} \textsf{true} if $u$ is validated (verified) by this process, else \textsf{false}
      \State $s_u \gets \mathsf{UNVERIFIED}$
      \State $\Sigma_u \gets 0$
      \State $\mathrm{EvUsers}_u \gets \emptyset$
      \ForAll{$r \in \Call{Shuffle}{\mathrm{Users}(\mathcal{C}_q) \setminus \{u\}}$}
        \State $\mathcal{T}_r \gets \{\, v \in \mathrm{Users}(\mathcal{C}_q) \setminus \{r, u\} \ : \ s_v=\mathsf{UNVERIFIED} \ \land\ r \notin \mathrm{EvUsers}_v \,\}$
        \State $\mathcal{T}_r \gets \mathcal{T}_r \cup \{u\}$
        \State $\mathcal{W} \gets \Call{CrossValQuery}{r,\ \mathcal{T}_r}$ \Comment{$\mathcal{W}$: users finalized this round}
        \If{$u \in \mathcal{W}$} \State \Return \textsf{true} \EndIf
      \EndFor
      \State \Return \textsf{false}
    \EndFunction
\medskip
    \Function{OnContactAdd}{$u$}
      \State \textbf{Precondition:} $u \in \mathrm{Users}(\mathcal{C}_q)$
      \State \textbf{Ensure:} \textsf{true} if $u$ is validated (verified) by this process, else \textsf{false}
      \State $b \gets \Call{OnKeyChange}{u}$ \Comment{$b$ is true if $u$ has been validated}
      \If{$s_u = \mathsf{VALID}$} \label{line:on-contact-add-if-start}
        \State $\mathcal{T} \gets \{\, v \in \mathrm{Users}(\mathcal{C}_q) \setminus \{u\} \ : \ s_v=\mathsf{UNVERIFIED} \,\}$
        \State \Call{CrossValQuery}{$u,\ \mathcal{T}$} \Comment{use $u$ as responder to help validate others}
      \EndIf \label{line:on-contact-add-if-end}
      \State \Return $b$
    \EndFunction
  \end{algorithmic}
\end{algorithm}

%% file: sections/appendix/dataset-analysis.tex
\section{Analyses of Social Datasets} \label{app:dataset-analysis}

\begin{table}[htbp]
    \centering
    \caption{Statistics of social datasets used in simulation. Average (\textit{Avg.}) metrics rounded to nearest integer. Friend counts computed only for users with public profiles.}
    \label{tab:dataset-statistics}
    \begin{tabular}{l|l|rrr}
        \toprule
       \multicolumn{2}{l|}{Metric} & \fbds & \pkds & \vkds \\
        \midrule
        \multicolumn{2}{l|}{\# of users} & \num{4039} & \num{1632803} & \num{50000} \\
        \multicolumn{2}{l|}{\# of users w/ public profiles} & \num{4039} & \num{1080278} & \num{23061} \\
        \hline
        \multirow{3}{*}{\makecell[l]{\# of friends \\ per user}} 
            & \textit{Avg.}   & \num{123} & \num{24} & \num{984} \\
            & \textit{Median} & \num{95}     & \num{12}    & \num{528}    \\
            & \textit{Q1}     & \num{49}     & \num{4}     & \num{205}    \\
        \hline
        \multirow{3}{*}{\makecell[l]{\# of cross-validating \\ friends}} 
            & \textit{Avg.}   & \num{48} & \num{3} & \num{25} \\
            & \textit{Median} & \num{41}    & \num{1}    & \num{6}     \\
            & \textit{Q1}     & \num{19}    & \num{0}    & \num{2}     \\
        \bottomrule
    \end{tabular}
\end{table}

In our simulation, we use three types of social datasets: \fbds, \pkds, and \vkds. The \fbds and \pkds datasets are from the \textit{Stanford Large Network Dataset Collection}~\cite{snapnets}. However, both were collected before 2012 and may not reflect the current structure of social networks. To address this limitation, we created the up-to-date \vkds dataset, collected from \vk, the largest social network in Russia. Table~\ref{tab:dataset-statistics} summarizes the dataset statistics. A ``\emph{cross-validating friend}'' in the metric ``\# of cross-validating friends'' is defined as follows: given an edge $(q, t) \in \mathcal{E}$, a querier's friend or responder $r$ is a \emph{cross-validating} friend for the edge if and only if $\left\lbrace(q, r), (r, t)\right\rbrace \subset \mathcal{E}$.

\paragraph{Data Collection Method}  
\fbds was collected by surveying 10 Facebook users, who manually labeled the social circles to which each of their friends belonged~\cite{FacebookPaper}. While the dataset includes friend relationships among all surveyed users, the collection method beyond the survey is not clearly described.  
\pkds was collected by crawling \pokec, the largest social network in Slovakia, using a breadth-first search (BFS) approach~\cite{PokecPaper}. The dataset includes all existing \pokec users as of 2011.  
For \vkds, we used the official \vk API~\cite{vksocialnetworkAPI2025} to crawl friend lists of public users. We attempted to crawl \num{50000} users, of which \num{23061} had public friend lists. Starting from a randomly selected user, we applied BFS to collect friend lists until reaching \num{50000} attempted users. Since the \vk API provides only mutual friend lists (not unidirectional follow lists), our dataset includes only mutual friendships.

\paragraph{Public/Private Profiles}  
\facebook allows users to set their friend lists as public or private, but~\cite{FacebookPaper} does not specify how private lists were handled. We therefore assume that \emph{all users in \fbds have public friend lists} (this may underestimate the performance of \sys but will not overestimate it).  
In \pkds, the public/private status of each profile's friend list is explicitly recorded. Similarly, \vk allows users to choose public or private settings. We observe significant proportions of private profiles in \pkds (33.8\%) and \vkds (53.9\%).  

While it is impossible to view the complete friend list of a private profile in \vk, if another public user is friends with a private profile, we can still observe that the two users are friends. Therefore, we collected friend lists of public users in \vk, and for private users, we constructed partial friend lists by including only public friends. Similarly, for \pkds, we observe that private profiles' friend relationships with public profiles are included in the dataset. Therefore, we employ the same approach as for \vkds.

For simulations using \pkds and \vkds, we used only public profiles as \emph{queriers}, since complete friend lists of private profiles cannot be retrieved. We include only public profiles as \emph{targets} as well. This is because if a target is private, we cannot count relationships where both the target and responder are private and friends, which severely degrades the simulation results. For \emph{responders}, we include both public and private profiles to fully utilize the available information (including partial friend lists of private profiles).

\paragraph{Relationship Tie Strength}  
The \fbds dataset reflects strong ties, as it was constructed by asking users to label their real-world social circles.  
\vk is also known to capture real-world relationships. However, given that over 50\% of users in our dataset have private profiles---which are more likely to represent real-world accounts with strong ties---the collected \vkds dataset may lean towards weaker ties relative to the actual \vk network. Similarly, public ``influencer accounts'' may have caused the dataset to overrepresent weaker ties.
According to~\cite{PokecPaper}, the most connected accounts on \pkds are commercial accounts, and the average user has approximately 22 friends. Considering that the overall average is \num{24}, the share of weak ties from commercial accounts is non-negligible. Furthermore, since \pokec supports unidirectional following, the dataset likely emphasizes weak ties more than \fbds.  

The ``\# of cross-validating friends'' in Table~\ref{tab:dataset-statistics} can be interpreted as an indicator of overall tie strength. As expected, we observe the ordering: $\fbds > \vkds > \pkds$.

%% file: sections/appendix/simulation-settings.tex
\section{Simulation Settings} \label{app:simulation-settings}

\subsection{Simulation Overview}

For each simulation iteration, we select a querier and simulate a situation where ``all the other users have already joined, and the querier is now joining the system and has just added all of their contacts.''
All others are assumed to have already completed key validations of their contacts (i.e., there are no \emph{honest but compromised} contacts).
The querier performs batch cross-validation to validate the public keys of all their contacts, and we record the number of SPRT iterations and queries required for each key validation.
We repeat this process for every user in the dataset as a querier.

For a querier $q$, a genuine or incorrect (malicious) public key is assigned to each of their contacts according to the server's behavior.
Let the contact $v$'s genuine public key be $pk_v$, and let the malicious public key (i.e., the key that the server wants the querier to accept) be $\mathit{pk}_{v}^{\mathsf{mal}}$.
Then the assigned key $\widehat{\mathit{pk}}_{v}^{q}$ is determined based on the following distribution:
\begin{equation}
    \Pr\left(\widehat{\mathit{pk}}_{v}^{q} = pk_v\right) = 1 - \sigmal
\end{equation}
\begin{equation}
    \Pr\left(\widehat{\mathit{pk}}_{v}^{q} = \mathit{pk}_{v}^{\mathsf{mal}}\right) = \sigmal
\end{equation}

\input{algorithms/simulation}

We simulate the worst-case scenario in which all the malicious responders in $q$'s contact list collude with the server. In this setting, $q$ receives the same value (i.e., $\mathit{pk}_{v}^{\mathsf{mal}}$) from both the server and all the malicious responders.
We use only public profiles in the datasets (i.e., users with public friend lists) as queriers and targets as mentioned in Appendix~\ref{app:dataset-analysis}. We use all users (both public and private) as responders.
The pseudocode for the simulation program is provided in Algorithm~\ref{alg:simulation}. We omit the cryptographic details from the original \sys algorithm and optimize it to evaluate a complete system involving over a million users.

\subsection{Parameter Selection}

This section describes the rationale for selecting the parameters $\alpha$, $\beta$, and $\sigmal$ used in the simulation.

\subsubsection{Selection of $\alpha$ and $\beta$}
We conducted preliminary simulations with various $\alpha$ and $\beta$ values to determine optimal settings. As a result, we found that a combination of $\alpha=10^{-3}$ and $\beta=10^{-2}$ yields a favorable balance between usability (i.e., high protocol success rate) and security (i.e., acceptable error rates).
Therefore, we use these values as defaults for our evaluation. Simulation results with different $\alpha$ and $\beta$ values are included in Appendix~\ref{app:alpha-beta-effects}.

\subsubsection{Selection of $\sigmal$}
Given our threat model in \S\ref{subsubsec:threat-model-server}, the server is assumed to be \emph{reputation-conscious} and unwilling to perform attacks if they can be easily detected.
Therefore, we set $\sigmal = 0.01$, implying that the server is expected to behave honestly in most cases.

\subsection{Effects of \texorpdfstring{$\alpha$ and $\beta$}{α and β}} \label{app:alpha-beta-effects}

\begin{table*}[htbp]
    \centering
    \caption{Simulation results of \sys on \vkds dataset with different $(\alpha, \beta)$ pairs. Parameters: $\mumal=0.05$, $\sigmal=0.01$.} \label{tab:alpha-beta-sensitivity}
    \begin{subtable}{0.7\linewidth}
        \centering
        \caption{Average number of SPRT iterations ($\nsprt$) and queries required for successful protocol completion. Query counts use batched ($n_{q}^{\mathrm{batched}}$) and unbatched ($n_{q}^{\mathrm{unbatched}}$) methods.}

        \begin{tabular}{l|l|rrr}
            \toprule
            \multirow{2}{*}{\makecell[l]{\textbf{Server status}}} & \multirow{2}{*}{\textbf{Metric}} & \multicolumn{3}{c}{$(\alpha,\beta)$}                                             \\
            \cline{3-5}
                                                                  &                                  & $(10^{-3},10^{-2})$                  & $(10^{-9},10^{-2})$ & $(10^{-3},10^{-9})$ \\
            \midrule
            \multirow{3}{*}{\makecell[l]{Honest                                                                                                                                         \\
            ($\widehat{\mathit{pk}_{(q,t)}}=pk_t$)}}                       & $\nsprt$                         & 2.22                                 & 2.22                & 8.87                \\
                                                                  & $n_{q}^{\text{unbatched}}$       & 81.23                                & 81.12               & 216.38              \\
                                                                  & $n_{q}^{\text{batched}}$         & 8.65                                 & 8.49                & 9.53                \\
            \hline
            \multirow{3}{*}{\makecell[l]{Malicious                                                                                                                                      \\
            ($\widehat{\mathit{pk}_{(q,t)}} = pk_t^{\mathsf{mal}}$)}}      & $\nsprt$                         & 3.31                                 & 8.84                & 3.31                \\
                                                                  & $n_{q}^{\text{unbatched}}$       & 111.37                               & 212.46              & 107.19              \\
                                                                  & $n_{q}^{\text{batched}}$         & 14.83                                & 32.48               & 4.34                \\
            \bottomrule
        \end{tabular}
    \end{subtable}
    \hfill
    \begin{subtable}{0.7\linewidth}
        \centering
        \caption{Average error and (protocol) failure rates across simulations. Observed error rates remain below expected thresholds ($\alpha, \beta$) in all cases.}

        \begin{tabular}{lrrr}
            \toprule
            \multirow{2}{*}{\textbf{Metric}}     & \multicolumn{3}{c}{$(\alpha,\beta)$}                                             \\
            \cline{2-4}
                                                 & $(10^{-3},10^{-2})$                  & $(10^{-9},10^{-2})$ & $(10^{-3},10^{-9})$ \\ \midrule
            False positive rate ($\hat{\alpha}$) & 0.0001                               & 0.0000              & 0.0001              \\
            False negative rate ($\hat{\beta}$)  & 0.0029                               & 0.0022              & 0.0000              \\
            \hline
            Failure rate (server honest)         & 0.0184                               & 0.0184              & 0.0781              \\
            Failure rate (server malicious)      & 0.0289                               & 0.0787              & 0.0291              \\
            \bottomrule
        \end{tabular}
    \end{subtable}
\end{table*}

Table~\ref{tab:alpha-beta-sensitivity} shows the simulation results of \sys with different $\alpha$ and $\beta$ values (with $\mumal=0.05$ and $\sigmal=0.01$). We observe that when $\alpha$ decreases (from $10^{-3}$ to $10^{-9}$), the number of SPRT iterations ($\nsprt$) and protocol failure rate increase significantly when the server is \emph{malicious}. Conversely, when $\beta$ decreases (from $10^{-2}$ to $10^{-9}$), $\nsprt$ and protocol failure rate increase significantly when the server is \emph{honest}. We can speculate that these metrics for both honest and malicious cases will decrease if we reduce $\alpha$ and $\beta$ simultaneously. These results suggest that error rates and usability of \sys exhibit a trade-off relationship, as discussed in \S\ref{subsubsec:alpha-beta-selection}.

%% file: algorithms/simulation.tex
\begin{algorithm}[htbp]
  \caption{Simulation Pseudocode of Cross-Validation}
  \label{alg:simulation}
  \begin{algorithmic}[1]
    \Require Directed contact graph $G=(\mathcal{V},\mathcal{E})$
    \Require Privacy predicate $\mathsf{Private}:\mathcal{V}\to\{\mathsf{true},\mathsf{false}\}$ \Comment{$\mathsf{Private}(u)$ is true iff $u$ has a private profile.}
    \Require True keys $\{pk_v : v\in\mathcal{V}\}$ and malicious substitutes $\{pk_v^{\mathsf{mal}} : v\in\mathcal{V}\}$
    \Require $\alpha,\beta, \mu_{\text{mal}}, \sigma_{\text{mal}} \in (0, 1)$
    \State \textbf{Enums: } $\mathsf{UNVERIFIED},\mathsf{VALID},\mathsf{INVALID}$; \textbf{evidence: } $e\in\{\Ematch,\Emismatch\}$
    \State \textbf{Notation: } $\Call{Bernoulli}{p}$ returns 1 with probability $p$ and 0 otherwise.
    \State \textbf{Notation: } $\mathrm{dom}(\mathbf M): \{\, t \in \mathcal{V} : t \mapsto a \in \mathbf M \,\}$ is the domain of map $\mathbf M$.

    \ForAll{$q \in \mathcal{V}$} \Comment{querier}
    \If{$\mathsf{Private}(q)=\mathsf{true}$} \textbf{continue} \EndIf
    \ForAll{$v \in \{\,v : (q,v)\in\mathcal{E}\,\}$} \Comment{contacts of $q$}
    \If{$\mathsf{Private}(v)=\mathsf{true}$} \textbf{continue} \EndIf
    \State $B_{(q,v)} \gets \Call{Bernoulli}{\sigma_{\text{mal}}}$  \Comment{server corrupt on edge $(q,v)$?}
    \State $\widehat{\mathit{pk}}_{v}^{q} \gets \begin{cases}
        pk_v^{\mathsf{mal}} & \text{if } B_{(q,v)}=1 \\
        pk_v                & \text{otherwise}
      \end{cases}$
    \State $\Sigma_{(q,v)} \gets 0$; \quad $\mathrm{KeyStatus}_{(q,v)} \gets \mathsf{UNVERIFIED}$
      \EndFor
    \ForAll{$r \in \{\,v : (q,v)\in\mathcal{E}\,\}$} \Comment{responders adjacent to $q$}
    \State $\mathcal{T} \gets \{\, t \in \mathcal{V} \setminus \{r\} \mid (q,t)\in\mathcal{E} \land \mathsf{Private}(t)=\mathsf{false} \land \mathrm{KeyStatus}_{(q,t)}=\mathsf{UNVERIFIED} \,\}$
    \State $M_{(q,r)} \gets \Call{Bernoulli}{\mu_{\text{mal}}}$ \Comment{$r$ malicious for this interaction?}
    \If{$M_{(q,r)}=0$} \Comment{honest responder}
    \State $\mathbf{A}_{(q,r)} \gets \{\, t \mapsto {\mathit{pk}}_{t} \;:\; (r,t)\in\mathcal{E} \,\}$
    \Else \Comment{malicious responder}
    \State $\mathbf{A}_{(q,r)} \gets \{\, t \mapsto pk_t^{\mathsf{mal}} \;:\; (r,t)\in\mathcal{E} \,\}$
    \EndIf
    \ForAll{$t \in \mathcal{T}$}
    \If{$t \in \mathrm{dom}(\mathbf{A}_{(q,r)})$}
    \State $a \gets \mathbf{A}_{(q,r)}[t]$ \Comment{key for $t$ advertised by $r$ to $q$}
    \State $e \gets \Ematch \ \mathbf{if}\ \widehat{\mathit{pk}}_{t}^{q} = a \ \mathbf{else}\ \Emismatch$
    \State $(\Sigma_{(q,t)},\, \delta) \gets \Call{SPRT}{\alpha,\beta,\mu_{\text{mal}},\, \Sigma_{(q,t)},\, e}$
    \If{$\delta \in \{\mathsf{VALID},\mathsf{INVALID}\}$}
    \State $\mathrm{KeyStatus}_{(q,t)} \gets \delta$
    \EndIf
    \EndIf
    \EndFor
    \EndFor
    \EndFor
  \end{algorithmic}
\end{algorithm}

%% file: sections/appendix/bandwidth-cost-eval.tex
\section{Bandwidth Cost Evaluation} \label{app:bandwidth-eval-details}

\begin{figure*}[htbp]
   \centering
   \begin{subfigure}{0.49\linewidth}
       \centering
       \includegraphics[width=0.8\linewidth]{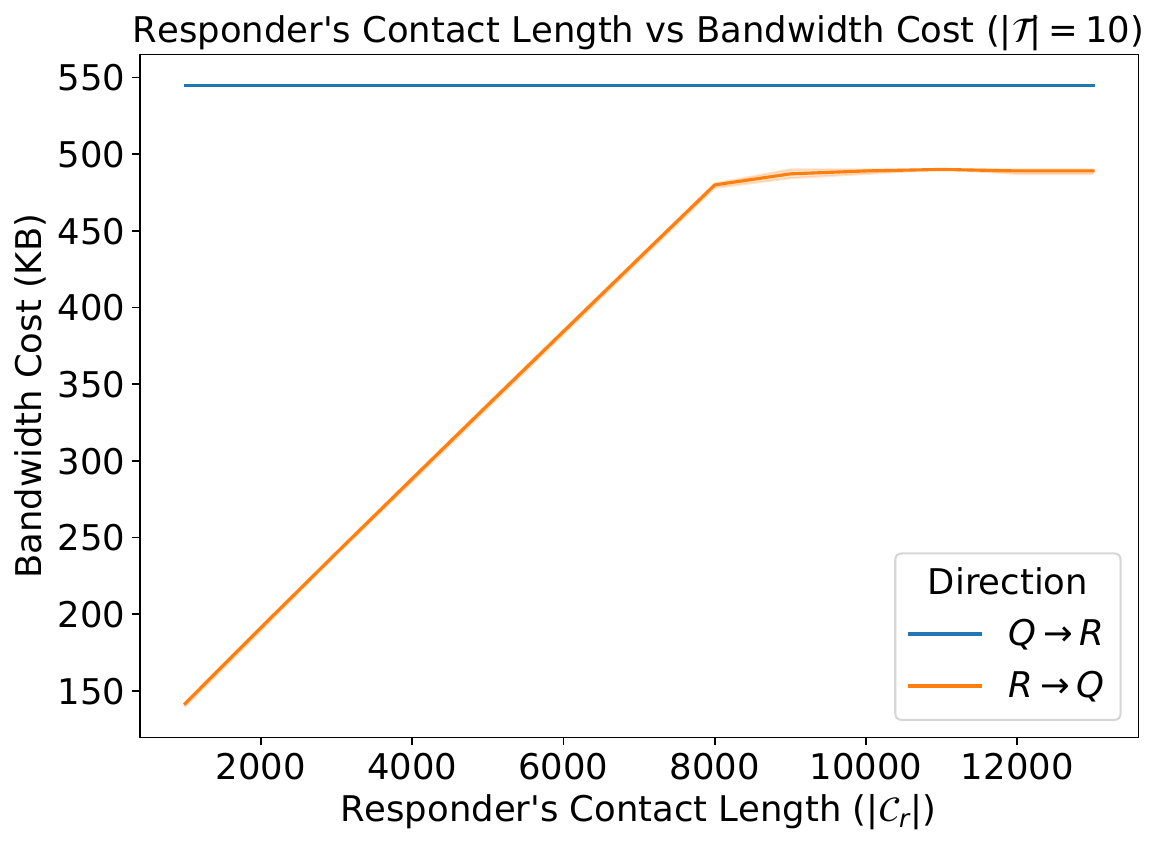}
       \caption{Bandwidth cost with varying responder contact list sizes ($|\mathcal{C}_r|$), with number of targets fixed at 10.}
       \label{fig:bandwidth-by-contact-list-size}
   \end{subfigure}
   \hfill
   \begin{subfigure}{0.49\linewidth}
       \centering
       \includegraphics[width=0.8\linewidth]{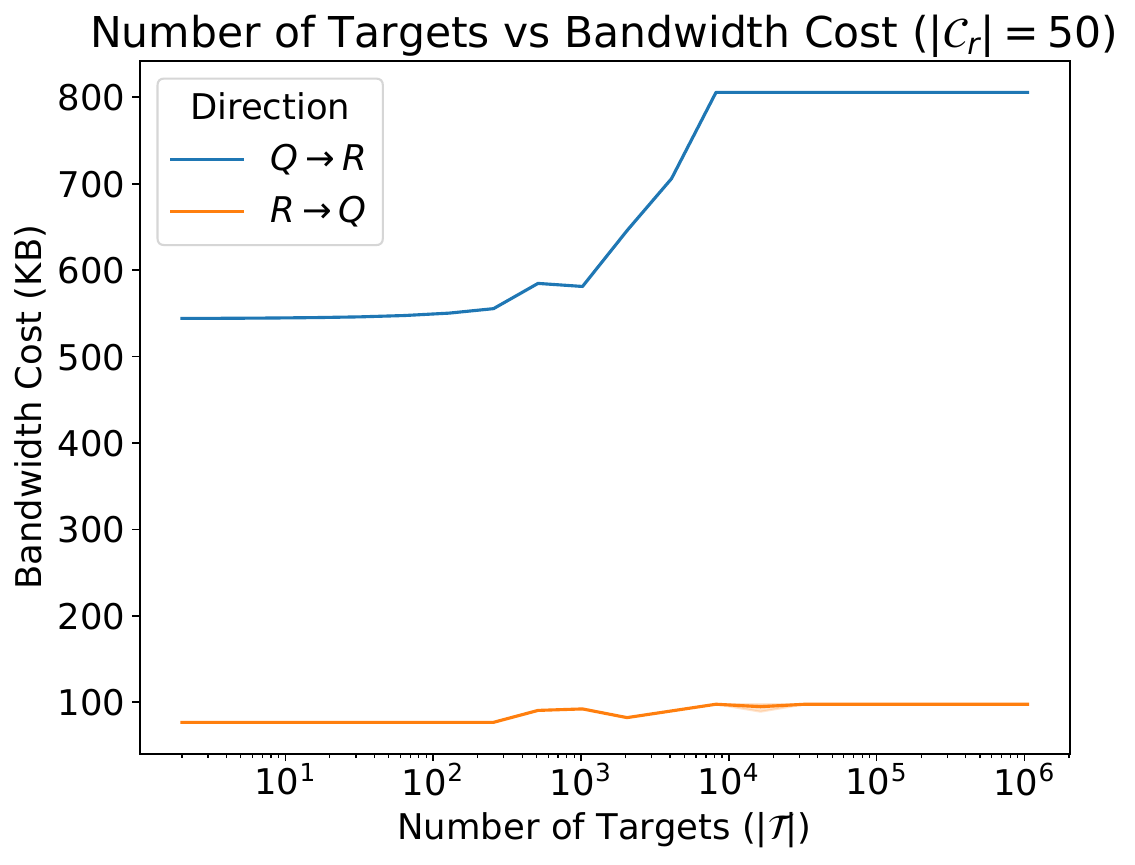}
       \caption{Bandwidth cost with varying number of targets ($|\mathcal{T}|$), with responder contact list size fixed at 50.}
       \label{fig:bandwidth-by-target-count}
   \end{subfigure}
   \caption{Bandwidth cost of \sys versus number of targets and responder contact list size. $Q \to R$ and $R \to Q$ represent querier-to-responder and responder-to-querier communication, respectively.}
   \label{fig:bandwidth-cost-summary}
\end{figure*}

Fig.~\ref{fig:bandwidth-cost-summary} shows the bandwidth cost of a \emph{single} execution of \proc{CrossValQuery} as we vary $|\mathcal{T}|$ (the number of targets) and $|\mathcal{C}_r|$ (the size of the responder's contact list).

Fig.~\ref{fig:bandwidth-by-contact-list-size} demonstrates that the responder-to-querier payload size grows linearly with $|\mathcal{C}_r|$ when $|\mathcal{C}_r| < 8000$. This is expected since the size of the encoded OKVS object is generally linear in the number of label-value pairs. Linear regression yields:
\[
y = 0.0521 \cdot |\mathcal{C}_r| + 81.76 \text{~KB},
\]
where $y$ is the total responder-to-querier payload size. In contrast, the querier-to-responder payload size remains constant around 557.79~KB, as $|\mathcal{C}_r|$ only affects the size of the OKVS object, which is transmitted in the opposite direction.

Fig.~\ref{fig:bandwidth-by-target-count} shows that the querier-to-responder bandwidth remains nearly constant at 550~KB when $|\mathcal{T}| < 1000$, increases linearly afterward, and plateaus beyond $|\mathcal{T}| > 10000$.
Since $|\mathcal{T}| \leq |\mathcal C_{q}|$, and $|\mathcal C_{q}|$ is usually small (e.g., fewer than 1000), the payload size can typically be considered constant. The responder-to-querier bandwidth remains nearly constant around 80~KB for typical users.

From these observations, the bandwidth cost of a \emph{single} \proc{CrossValQuery} execution can be approximated as:
\begin{itemize}
    \item Querier to responder: 550~KB
    \item Responder to querier: $0.0521 \cdot |\mathcal{C}_r| + 81.76$~KB
\end{itemize}

